\documentclass[11pt, preprint]{aastex}

\shorttitle{Mid-Infrared Imaging of Starburst Galaxies}
\shortauthors{Soifer  et al.}

\begin{document}

\title{High Resolution Mid-Infrared Imaging of Infrared Luminous Starburst
Galaxies
\altaffilmark{1}
}

\author{B. T. Soifer\altaffilmark{2}, G. Neugebauer, K. Matthews, E.
Egami, A. J. Weinberger\altaffilmark{3}}
\affil{Palomar Observatory, California Institute of
Technology, 320-47, Pasadena, CA 91125}
\email { bts@mop.caltech.edu, gxn@mop.caltech.edu, kym@caltech.edu,
egami@mop.caltech.edu,alycia@astro.ucla.edu}
\author {M. Ressler}
\affil {Jet Propulsion Lab, 169-506, 4800 Oak Grove Dr., Pasadena, CA 91109}
\email {ressler@cougar.jpl.nasa.gov}
\author {N.Z. Scoville}
\affil {Divison of Physics, Mathematics and Astronomy, California Institute of
Technology, 105-24, Pasadena, CA 91125}
\email {nzs@astro.caltech.edu}
\author {S. R. Stolovy}
\affil {SIRTF Science Center, California Institute of Technology, 
314-6, Pasadena, CA 91125}
\email {stolovy@ipac.caltech.edu}
\author { J.J. Condon}
\affil {National Radio Astronomy Observatory, 520 Edgemont Road,
Charlottesville, VA 22903}
\email { jcondon@nrao.edu}
\author{ E.E. Becklin}
\affil{Department of Physics and Astronomy, UCLA, Los Angeles, CA 90095}
\email {becklin@astro.ucla.edu,}

\altaffiltext{1} {Based in part on observations obtained at the
W. M. Keck Observatory  which is operated as a scientific partnership
among the  California Institute of Technology, the  University of
California and the National Aeronautics and Space Administration.}

\altaffiltext{2} { Also at SIRTF Science Center, California Institute of
Technology, 314-6, Pasadena, CA 91125}

\altaffiltext{3} { Currently at Department of Physics and Astronomy,
University of California Los Angeles, 156205 Los~Angeles, CA 90095}


\begin{abstract}

Observations for seven infrared luminous starburst galaxies  are
reported in the mid-infrared from 8 - 18~$\mu$m using the Keck
Telescopes with  spatial resolution approaching the diffraction limit
.  All of the galaxies observed show evidence of strong interactions
based on optical morphologies.  For these galaxies, a substantial
fraction, usually more than 50\%,  of the infrared luminosity is
generated in regions ranging in sizes from 100pc -- 1 Kpc.  Nuclear
starbursts often dominate the infrared luminosity, but this is not
always true.  In some galaxies, most notably NGC 6090, substantial
infrared luminosity greatly in excess of the nuclear luminosity is
generated in regions associated with the physical interaction between
two galaxies. The radio emission is a good tracer of the location of
high luminosity young stars.

The visual/ultraviolet  radiation output of the nearby
star forming galaxies is dominated by emission from regions that are
generally not producing the copious infrared luminosity of the
systems.  As seen in comparing the mid-infrared and near infrared
images of the galaxies observed here,  the regions of high infrared
luminosity in local galaxies  are significantly smaller than the
galaxies as a whole.  The integrated spectral energy distributions
(SEDs)  of these galaxies are very different from the SEDs of the
regions of  star formation. If the SEDs of star-forming regions in
these galaxies reflect the SEDs that would be found in forming
galaxies at high redshift, we would expect the distant galaxies to be
dominated  by the mid and far-infrared luminosity output far more
than the integrated luminous output of nearby starburst galaxies would
suggest.

\end{abstract}

\newpage

\section{Introduction}

Starburst galaxies have long been known to be copious emitters of
infrared luminosity.  From early observations of nearby starburst
systems such as M82 and NGC 253 (see e.g. Kleinmann \& Low, 1970,
Rieke \& Low, 1972, Harper \& Low, 1973), it has been clear that the
bulk of the energy of these systems emerges in the thermal infrared.
Analysis of the IRAS Bright Galaxy Sample showed that 30\% of the
energy output in the local universe emerges in the mid and far infrared (5
-- 200~$\mu$m) (Soifer and Neugebauer, 1990) and the bulk of this
luminosity is due to star formation in these galaxies.  Recently, the
idea that the bulk of the radiant energy emitted in the universe has
been emitted at infrared wavelengths has been raised through analysis
of the far infrared background as measured by the COBE satellite
(Hauser et al. 1998, Puget et al. 1996).  Further, deep submillimeter
surveys have suggested that much of this luminosity is being generated
in forming galaxies at redshifts of z$\geq$ 2  (e.g., Eales et
al. 2000).

Infrared luminous starbursts in the local universe afford us the best
opportunity to study the processes by which such systems form and
evolve, and provide detailed examples where we can test physical
models of such systems.  The  IRAS all sky survey (Joint IRAS Science
Team, 1989) has provided a unique database from which many such
systems have been chosen for detailed examination.

A wide variety of observational techniques have been employed to probe
infrared luminous starburst galaxies, ranging from optical
spectroscopy (e.g., Kim et al. 1995, Veilleux et al. 1995),
mid-infrared imaging and spectroscopy (Keto et al. 1997, Miles et
al. 1996, Roche et al.  1991, Dudley 1999) to cm radio continuum
imaging (Condon et al. 1990, 1991, hereafter C91). The nuclear
environments of starburst galaxies have been effectively imaged with
high spatial resolution in the  near infrared by e.g. Scoville et
al. (2000, hereafter S00), Dinshaw et al. (1999) and Alfonso--Herrero
et al. (2000a,b).

Because dust responds through its thermal emission instanteously to
the radiation field in which it is embedded, thermal radiation by dust
is the ideal way to trace the current location of the dominant
luminosity sources, i.e. the most luminous stars, in starbursts.  In
very dusty systems, such as those known  from IRAS observations to be
luminous starbursts, high spatial resolution imaging in the thermal
infrared  can address the location and extent of the current star
formation in these galaxies unimpeded by foreground extinction.   
How the thermal dust emission is related to the other
tracers of star-formation is important for utilizing these tracers in
more distant galaxies.

IRAS measured the total bolometric luminosity from these galaxies
(e.g. Soifer et al. 1987).  The spatial resolution of those
observations,  $\sim 1'-2'$, was, however, inadequate to locate the
emission within the galaxies.  Because the mid-infrared wavelengths
(8--25~$\mu$m) carry a significant fraction of the total bolometric
luminosity in infrared bright galaxies,  ranging from $\sim$10\% to
$>$ 30\% of the 8--1000~$\mu$m luminosity, observations at these
wavelengths with a resolution $\le 1''$  have the potential to
directly trace the locations where  the most luminous stars are
forming in these systems.  With a diffraction limit  of  0.24$''$ FWHM
at 10~$\mu$m, the Keck Telescope provides a substantial improvement in
spatial resolution over previous mid-infrared observations,  probing
the distribution of the thermal emission  at the 100--300 pc scale in
luminous starbursts at distances of 40--120 Mpc.

In this paper we report imaging observations from 8 -- 18~$\mu$m of a
sample of highly luminous starburst galaxies  at spatial resolutions
of 0.3-0.6$''$.  These data provide  the highest spatial resolution
yet achieved of the thermal emission from these systems  and trace the
spatial distribution of the emergent luminosity in these systems.
For purposes of establishing luminosities and physical distances in
these galaxies we adopt H$_o$=75~km~s$^{-1}$Mpc$^{-1}$.

\section {The Sample}

The objects observed were taken from the IRAS Bright Galaxy Sample
(BGS, Soifer et al. 1987, 1989)  The basic information for the seven
objects in the sample is given in Table 1.  Since the goal of this
study  is to understand  luminous starburst galaxies, systems having
Seyfert like spectra were excluded.  With this selection criterion,
there are 25 objects in the BGS in  the luminosity range 11.4 $\le$
log(L$_{\it bol}$[L$_{\odot}$])$\le$ 11.9; seven were observed. 
These seven objects
are among the closest highly infrared luminous starbusts galaxies.  
Six of the seven objects observed have clear
evidence of strong interactions/mergers from optical images. The only
object without such evidence, Markarian 331=UGC 12812, has several
close companions.  In this luminosity range 60-70\% of objects are
found in strongly interacting or merging systems (Sanders et al. 1988,
Sanders and Mirabel, 1996).   The objects were
selected to be detectable from the ground at 12~$\mu$m, so that there
is a  bias towards brighter 12~$\mu$m flux densities in these
galaxies, as measured by IRAS, compared to the galaxies in the same
luminosity range in the BGS. Three of the galaxies in the sample, IC
1623=VV114, NGC 3690/IC 694=Arp 299 and NGC 1614=Arp 186 are the three
brightest objects at 12~$\mu$m in the above sample.  This selection of
brighter 12~$\mu$m flux density appears to be the major bias in our
sample as compared to the total BGS sample for this luminosity range.

\section {Observations and Data Reduction}

The observations were made  using the MIRLIN mid-infrared camera
(Ressler et al. 1994) at the f/40 bent Cassegrain visitor port of the
Keck II Telescope, and the imaging mode of the Long Wavelength
Spectrograph (LWS, Jones and Puetter, 1993) at the f/25 forward
Cassegrain focus of the Keck I Telescope.   In general, only
observations where the FWHM of the PSF  was less than 1.0$''$ were
used. Because of the variability, the seeing conditions are described
throughout the text when specific objects are discussed.

The MIRLIN camera uses a 128$\times$128 Si:As array with a plate scale
of 0.138$''$/pixel for a total field of view of 17$'' \times$17$''$.
At each wavelength the observing procedure was the same.  A secondary
with a square wave chop of  amplitude 6$''$ in the north-south  or
east-west direction at 4 Hz was employed for fast beam switching.  The
frames sampling  each chop position were coadded separately in
hardware, resulting in two images.  After an interval of approximately
a minute, the telescope was nodded perpendicular to the chop direction
(east-west or north-south) by 6$''$ and a second pair of images was
obtained in order to cancel residuals in the sky and to subtract
telescope emission.  This procedure was repeated a number of times at
each  wavelength.  The data were reduced by differencing the two
images obtained within the chop pairs at each nod location. Then
the resulting positive images were coadded, with the positions
appropriately adjusted to a common location, to yield an image
centered in a field approximately 6$'' \times$6$''$.  Because of the
chopper and telescope nod spacings employed for the observations, the
data are not capable of measuring low surface brightness emission that
is extended beyond a 6$''$ diameter region. In addition extended emission
within a 6$''$ region with a surface brightness of 5 mJy/square arcsec is 
not detectable within these images.

Observations with MIRLIN of VV114, NGC 1614, NGC 3690 and  Markarian
331 were obtained in March and October 1998. The MIRLIN observations
were made under photometric conditions.  The observations of the
targets were interleaved with  observations of nearby bright stars
that served as photometric calibration and  to establish the point
spread function (PSF) for the observations.

Observations of all the galaxies listed in Table 1 were made with LWS
in October 1999, and January, May and September 2000.  Generally these
observations were made under photometric conditions. Over the course
of the observations, the mid-infrared seeing varied from near
diffraction limit($\sim 0.3''$ at 12.5$\mu$m) to $> 1.0''$. 

The secondary chopper was set to an amplitude of 5$''$ at a  
frequency $\sim$ 5 Hz for all the objects except NGC~1614, NGC~2623, IC~883,
where the chopping amplitude was 4$''$. Observations were made  in a
fashion similar to  the MIRLIN observations, except that the telescope
nodding was in  the same direction as the chopping and the nodding
amplitude was the same as the chopping amplitude. The data were
reduced in a manner similar to the MIRLIN data.  The pixel scale is
0.08$''$ pixel$^{-1}$.

The filters defining the photometric bands are quite similar in both
MIRLIN and LWS and all have widths about 10~\% of the central
wavelengths which are listed in Table~2. Details of the filters are
given in Soifer et al (1999).  The photometry  was calibrated based on
observations of four bright stars, $\alpha$ Tau=HR~1457
([12.5~$\mu$m]=-3.07~mag),  $\alpha$ Boo=HR~5340
([12.5~$\mu$m]=-3.15~mag), $\beta$ Peg=HR~8775
([12.5~$\mu$m]=-2.55~mag),  and $\alpha$ Cet=HR~0911
([12.5~$\mu$m]=-1.92~mag) whose magnitudes, in turn, were based on
IRAS and Keck data. The magnitudes  adopted at the other mid-infrared
wavelengths for these stars were  within 0.08~mag of these values. The
uncertainties in the photometry, based on the internal consistency of
the observations, is estimated to be 5\% at $\lambda \le$17.9~$\mu$m.
The flux density corresponding to 0.0 mag (Vega-based)  was taken to
follow the prescription given in the Explanatory Supplement to the
IRAS Catalogs and Atlases (Beichman et al. eds, 1989), and is given in
Soifer et al. (1999).

In addition, high spatial resolution observations of two of these
galaxies, VV114=Arp 236 and NGC 3690=Arp 299 were obtained at
3.2~$\mu$m with NIRC (Matthews and Soifer, 1994) on the Keck I
telescope in Dec 1996. Although the background is higher at this
wavelength than at shorter wavelengths, observations were made in the
standard ``stare and dither'' mode for near infrared observations.
Observations were obtained for Markarian~331 at 2.15~$\mu$m(K$_s$) and
3.4~$\mu$m (L$'$) using the near infrared camera on the 200-inch Hale
Telescope. Observations were made in a similar fashion to those on the
Keck Telescope.

\section {Results and Discussion for Individual Galaxies}

The basic observational results are presented in Table~2 and Figures
1--7. The observed flux densities, measured from the Keck images in
the largest beams feasible for each of the galaxies, are presented  in
Table~2. All of the galaxies were observed at 11.7 and 12.5~$\mu$m,
while most of them were observed at 17.9~$\mu$m.  For comparison, the
IRAS flux densities at 12$\mu$m, representing the integrated flux
densities for the galaxies, are also presented in Tabel~2.

In the images, the locations of the origin and in some cases  other
positions, are marked in each image to facilitate comparisons.
Because bright stars are not common in the  mid-infrared, the {\it a
priori} astrometric registration of the mid-infrared images with
respect to other images is not better than 1 $''$. The locations of
the images in these figures were determined by identifying
morphologically similar features and assuming that they are spatially
coincident at wavelengths between the radio and near infrared. In the
mid-infrared  morphological features within $\sim 5''$ diameters were
imaged on a single array and thus differences on that scale between
images at different wavelengths are robust.

In the plots that present the spectral energy distributions (SEDs) for
these  galaxies, a variety of data are presented.  The integrated flux
densities at each wavelength are from 2MASS and IRAS data. The large
beam Keck data are presented (from Table 2) as well as data from
NICMOS (or groundbased images) with the photometry scaled to the same
size beams.  In addition, to illustrate how  the SEDs vary with beam
size, we also present multiaperture photometry at the locations where
there is significant 12.5$\mu$m emission in the Keck images.  The beam
diameters range from 1$''$ to 4$''$, depending on the size and
complexity of the source.  Photometry is taken from Keck mid-infrared,
NICMOS and groundbased near  infrared data. The details of the small
beam photometry are explained in the corresponding figure captions.

In the following sections we discuss the observational results for the
galaxies separately.

\subsection {VV 114=IC1623=Arp 236}

At a redshift of cz=6000 km/s (80 Mpc, 400 pc/$''$) VV114=Arp 236 is a
highly disturbed system that shows two major centers in optical
images, with low surface brightness tails evident over 55$''$ (22 Kpc)
in photographic images (Arp 1966). The two centers are aligned
east-west and separated by about 20$''$ (8 Kpc). The bolometric
luminosity of the system is $\sim 4 \times 10^{11}L_{\odot}$ (Soifer
et al. 1987), more than  90\% of which emerges at far infrared
wavelengths.

Previous detailed near infrared studies of VV114  have been reported
by Knop et al. (1994) and Doyon et al. (1995).  High resolution near
infrared imaging of VV114 has been reported by  S00.  Radio imaging of
this galaxy at 1.4~GHz has been presented by Condon et al. (1990),
and at 8.4~GHz  by C91.  CO observations of VV114 were reported by Yun,
Scoville and Knop (1994), while submillimeter continuum imaging of
this system is presented in Frayer et al. (1999).

The images presented in Knop et al. (1994), ranging in wavelengths from
B(0.44~$\mu$m) to L(3.5~$\mu$m), show the core of the eastern galaxy is
nearly invisible at B and becomes brighter with respect to the western
nucleus at longer wavelengths. In the longest wavelength images of Knop
et al.  the eastern core is clearly double, with a separation
between peaks of $\sim$2$''$ (800 pc).  In the eastern system
the southwestern component becomes increasingly dominant at longer
wavelengths. In the L band image of Knop et al. the southwestern
component is the only one detected.  ISO imaging of this galaxy at 7 $\mu$m
(Charamandlis et al 1999) shows a bright peak apparently associated
with the eastern nucleus with an extension to the west.

The only portion of the galaxy detected in the MIRLIN and LWS imaging
was the eastern source.  The 12.5~$\mu$m image of this nucleus is
presented in Figure~1a, along with  the 3.2~$\mu$m image obtained with
NIRC, the 2.2~$\mu$m  NICMOS image from S00 and the 8.4~GHz  image of
C91. The 12.5~$\mu$m image presented in Figure~1a is smoothed with a
Gaussian with  half the diffraction limited FWHM  to improve the
signal to noise ratio. The effective resolution of this image is
0.34$''$, i.e. 10\% greater than the formal diffraction limit at
12.5~$\mu$m.

The two peaks in the eastern nucleus are clearly seen at 12.5~$\mu$m;
both components are resolved in the image. In the subsequent
discussion they are referred to as VV114E$_{NE}$ and VV114E$_{SW}$
respectively. VV114E$_{NE}$ is centrally concentrated with an apparent
size of 0.44$''$.  The VV114E$_{SW}$ component is complex, with a main
northern component having an apparent size of 0.51$''$ and several
sources to the south, separated from the brighter peak by 0.6$''$ to
0.8$''$.  The peak positions of VV114E$_{NE}$ and VV114E$_{SW}$, as determined
in the NICMOS 2.2~$\mu$m image, are marked in all four images of
Figure~1a.  The marks are located at the same relative positions in 
all four panels of the montage.

The 3.2~$\mu$m image, with lower angular resolution, shows structure
that appears quite similar to that at 12.5~$\mu$m.  At 3.2~$\mu$m
VV114E$_{SW}$ is only marginally resolved; while VV114E$_{NE}$ is
approximately 0.7$''$ in diameter and apparently at the same
orientation as the source seen at 2.2$\mu$m.  At 3.2~$\mu$m VV114E$_{SW}$
appears displaced from the corresponding  12.5~$\mu$m source by
0.1$''$.  The radio image shown in Figure 1a shows a general
similarity to the 12.5~$\mu$m image with a bright isolated northeast
component, and a southwest source that consists of several distinct
components separated by $\sim$0.4--0.8$''$. The radio map of VV114E$_{SW}$
shows that the components are reasonably close in peak brightness,
like the 12.5~$\mu$m image.

At 2.2~$\mu$m the structure of VV114E is significantly different from
the mid-infrared and radio structure.  VV114E$_{NE}$ shows a significant
ellipticity at 2.2~$\mu$m, with a FWHM size of 0.8$'' \times$ 0.4$''$
at a position angle of 60~\degr, suggestive of a stellar nucleus.
VV114E$_{SW}$ breaks up into two sources, a bright, apparently unresolved
source nearly  coincident in position with the 12.5~$\mu$m source, and
a fainter source 0.86$''$ to the southwest of the brighter point-like
source.

While consistent in overall structure, the images do not agree in detail at the different wavelengths.    The
registration between images at different wavelengths is based on the
{\it assumption} that VV114E$_{NE}$ is spatially coincident at all
wavelengths. Since this source is reasonably compact and isolated from
other sources by $\sim$ 2$''$ at all wavelengths, we believe that this
is a reasonable assumption. 

At 2.2~$\mu$m and 3.2~$\mu$m the positions of VV114E$_{NE}$ and VV114E$_{SW}$
agree in position angle to $<$1~\degr and in separation to within
0.03$''$. In addition, as noted above, at  3.2~$\mu$m the northeastern
source appears slightly elongated in the same sense as the elliptical
shape of this source at 2.2~$\mu$m. There is also a slight  distortion
of the lower level contours of the southwestern  source at 3.2~$\mu$m
consistent with a fainter source at the location of the faint source
to the southwest of the point source in the 2.2~$\mu$m image.  Thus
given  the angular resolution and signal to noise ratio in the
3.2~$\mu$m image,  these images appear the same.

The 12.5~$\mu$m and 2.2~$\mu$m images differ in detail. In  both cases
the  northeastern source is extended, and there is some suggestion of
extended emission at 12.5~$\mu$m at a similar position angle as the
major axis of the 2.2~$\mu$m image.  There is a significant
disagreement in position and morphology of the southwestern source
between these wavelengths.  This is illustrated in Figure~1b, where
the 12.5~$\mu$m contours are overlaid on a grayscale version of the
2.2~$\mu$m  image.  The northeast and southwest peaks in the two
images appear at the same position angle of 80~\degr  but the location
of the brightest component of the southwestern  source disagrees by
0.1$''$ between the two wavelengths.  Furthermore, the secondary
peaks to the south and southwest  of the southwestern  peak at
12.5~$\mu$m do not appear to have any corresponding peaks in the
2.2~$\mu$m image, while the faint peak to the southwest of the bright
2.2~$\mu$m peak  has no counterpart in the 12.5~$\mu$m image.

The largest discrepancies in location of the peaks are between the
12.5~$\mu$m image and the 8.4 GHz image.  As can be seen in Figure 1a,
the separations between the VV114E$_{NE}$ and VV114E$_{SW}$ peaks differ by
0.3$''$ or 100 pc. There is no {\it a priori} means of knowing whether
the registration we have chosen (forcing VV114E$_{NE}$ to coincide at both
wavelengths) is valid, or whether forcing the peak of VV114E$_{SW}$ to
coincide might be appropriate. In either case, a discrepancy
exists between the locations of the mid-infrared and radio peaks.

Figure~1b also shows the 12.5~$\mu$m and the 3.2~$\mu$m contours
overlaid  separately on a grayscale version of the 8.4~GHz image.   In
both cases the location of the peak of the southwestern   source in the
``thermal" infrared does not coincide with the radio peak.  As in the
case of comparing the sources at 12.5~$\mu$m and 2.2~$\mu$m, the
northeast and southwest   peaks fall along the same position angle, but
the separations  disagree.  The separation of the radio peaks is
1.35$''$, compared to 1.55$''$ in the  thermal infrared. The structure
to the south of the southwestern  peak appears to agree much  better
between the radio and mid-infrared images, showing agreement in both
location and elongation.

In summary, {\it assuming} VV114E$_{NE}$ is spatially coincident at all
wavelengths, the VV114E$_{SW}$    peaks coincide at 2.2 and 3.2~$\mu$m, and
are displaced from the 12.5~$\mu$m peak by 0.1$''$ (30 pc).  The
12.5~$\mu$m peak is displaced from the 8.4GHz  peak by 0.2$''$(80pc);
at  12.5~$\mu$m VV114E$_{SW}$    lies between the 8.4~GHz and 2.2/3.2~$\mu$m
peaks, and they all lie on the same line from VV114E$_{NE}$.

The flux density of VV114 measured by IRAS at 12 $\mu$m is 0.8 Jy in
an unresolved beam and 1.1 Jy in total.  The flux density measured in
the imaging of VV114E is 0.34 Jy at 12.5~$\mu$m, significantly less
than the IRAS values. To attempt to detect additional emission,  the
position of peak extended infrared emission in VV114W based on the ISO
image of Laurent et al.  (2000) was imaged as well.  This location is
14$''$ west of the VV114E position.  No flux was detected in the Keck
image at 12.5 $\mu$m, with a limit of 8$\pm$5 mJy in a 4$''$ diameter
beam. For this imaging the chopping amplitude was 10$''$. The
non-detection of the western source places an upper limit of
$\sim$1mJy/square $''$  on the surface brightness of the emission at
this location.

The western galaxy in the VV114 pair is detected at 3.2~$\mu$m, and is
shown in Figure~1c.  The full field containing both systems is shown
at 2.2~$\mu$m and 3.2~$\mu$m, where the 3.2~$\mu$m contours are
overlaid on the 2.2~$\mu$m grayscale. The 3.2~$\mu$m NIRC image shows
that there is low surface brightness, extended emission associated
with the western galaxy and the interaction region between the two
galaxies.  The 3.2~$\mu$m emission is comprised of a combination of
photospheric, nebular and dust emission.  The 3.2~$\mu$m filter
includes the 3.3~$\mu$m feature commonly identified as due to
Polycyclic Aromatic Hydrocarbons, or PAHs (e.g., Dale et al. 2000) and
so there is a contribution from this mechanism as well.  The detection
of extended emission at 3.2~$\mu$m, in addition to the  detection of
substantial extended emission in the IRAS photometry with a $\sim$1$'$
beam,  argues that the non-detection of the majority of the IRAS
12~$\mu$m flux is probably the result of the difficulty in detecting
low surface brightness extended emission in the thermal infrared.

The spectral energy distributions for the sources in VV114 from 1.2 to
25$\mu$m are presented in Figure 1d as flux per octave ($\nu
f_{\nu}$).  In addition to the total fluxes from IRAS and 2MASS,
photometry in a 4$''$ diameter beam centered on VV114E, as well as in
1$''$ diameter beams centered on VV114E$_{NE}$ and VV114E$_{SW}$, are
presented to illustrate how the SEDs change with observing beam. The
small beam photometry at 2.2$\mu$m was scaled from the NICMOS image in
S00, while the 3.2$\mu$m data are from our NIRC imaging. The lower
panel illustrates the photometry at the corresponding location in
VV114W.

Both VV114E$_{NE}$ and VV114E$_{SW}$ show a significant drop in flux at
10~$\mu$m as compared to the observed flux at 8 and 12~$\mu$m.  This
is generally attributed to absorption by cold silicate dust overlaying
warmer emission.  A substantial uncertainty in quantifying the
overlaying absorption arises from the contribution to the underlying
emission from the aromatic (PAH) features.  The spectrum  of VV114
reported by Dudley (1999) with a 5.5$''$ diameter beam includes all of
the emission seen in the image in Figure~1a.  This spectrum shows  an
emission feature at 11.3~$\mu$m (rest wavelength)  attributed to  PAH
emission. The spectrum of Dudley, if smoothed to the resolution of the
filters used for our imaging, is consistent with the apparent ratio of
fluxes at 11.7 and 12.5~$\mu$m as shown for the flux from VV114E in
the large beam shown in Figure~1d.

The separate spectral energy distributions (SEDs) of VV114E$_{NE}$ and
VV114E$_{SW}$ differ significantly. The decrease in flux density at
10~$\mu$m of VV114E$_{NE}$ is substantially deeper than in
VV114E$_{SW}$.  The composite mid-infrared spectrum of galaxies
presented by Dale et al. (2000), if representative of the underlying
emission spectrum for VV114E, suggests that the extinction of an
intrinsic PAH like spectrum would lead to an absorption optical depth
of about 0.5 at 10~$\mu$m in VV114E$_{SW}$.  This is also consistent
with the fact that the southwest source appears to dominate the VV114E
flux throughout most of the 10~$\mu$m wavelength range. VV114E$_{NE}$
appears to have an optical depth $\tau$ in the range 1--2
greater than that of VV114E$_{SW}$ at 9.7~$\mu$m (i.e. a net
differential optical depth, $\tau$, of 1.5--2.5 at
$\sim$10~$\mu$m). The absorption optical depths cited here are by
comparison to zero absorption at 8 or 12 $\mu$m.

Yet another complication is the fact that the dust optical depth is
not zero at 8 and 12~$\mu$m, as assumed above.  Li and Draine (2000)
have recently compiled  the properties of interstellar
dust opacity (predominantly due to silicate absorption) 
that would result in an increase of a factor of 1.3 to
the optical depths derived above to take into account the extinction
at 8 and 12~$\mu$m.

Thus we infer that the silicate optical depths are $\sim$ 0.7  and
2--3 for VV114E$_{SW}$ and VV114E$_{NE}$ respectively.  These results, which
imply VV114E$_{NE}$ is more heavily obscured than VV114E$_{SW}$, contradict
the extinctions derived from the near infrared imaging of these
sources. As can be seen directly from Figure~1 of Knop et al. (1994),
VV114E$_{SW}$ is significantly ``redder'' than VV114 NE, i.e., it becomes
significantly brighter than VV114~NE at 2.2~$\mu$m compared to 0.8
$\mu$m. To the extent that this reflects overlaying extinction, it
suggests substantially more extinction obscuring VV114E$_{SW}$ than
VV114E$_{NE}$ .

\subsection {NGC 1614=Arp 186}

NGC 1614 is a strongly interacting galaxy at a redshift of cz=4800
km/s (64 Mpc, 320 pc/$''$).  The optical photograph of Arp (1966)
shows crossed  tails with a total extent of $\sim$30 Kpc.  Its
infrared luminosity is $4 \times 10^{11}L_{\odot}$ (Soifer et
al. 1987).  A detailed visible,  infrared and radio study by Neff et
al. (1990) shows an HII optical spectrum and tidal tails. The radio
image at 5~GHz  shows a ring $\sim$ 1.2$''$ (380 pc) in diameter
(Neff et al.). High resolution near infrared imaging and spectroscopy
(Alonso-Herrero et al. 2000b) shows a starburst nucleus of
$\sim$0.3$''$ (100pc) diameter   revealed through strong photospheric
CO absorption in supergiants.   This nucleus is surrounded by a ring
of very large HII regions. The ring diameter of $\sim$1.2$''$(380 pc)
is traced in the P$\alpha$ image (Alonso-Herrero et al.).  Miles et
al. (1996) have reported imaging of NGC 1614 at 11.7~$\mu$m with 1$''$
resolution.

Figure~2a shows the 12.5~$\mu$m contour map of NGC 1614, the 4.8~GHz 
radio contours of Neff et al (1990), the 2.2~$\mu$m broadband and the
1.87~$\mu$m P$\alpha$ NICMOS images from Alonso-Herrero et al.(2000b).
The 12.5~$\mu$m image, obtained in excellent (diffraction limited)
0.30$''$ seeing, shows a ring-like structure of diameter $\sim$1.2$''$
and overall extent of $\sim$1.7$''$ that appears virtually identical
in overall appearance to the ring seen in the radio at 4.8~GHz  and in
P$\alpha$.

A detailed comparison of the 12.5~$\mu$m and the P$\alpha$ and 4.8~GHz 
images is presented in  Figure~2b, where the contours of the
mid-infrared image are overlaid on  the grayscale images of P$\alpha$
and 4.8~GHz  radio emission.  The images were superimposed by matching
by eye the centroids at the different wavelengths.  The peaks in the thermal
infrared, P$\alpha$ and 4.8~GHz  images are very well matched,
particularly to the  southeast where the peaks agree in size and
orientation.

In contrast to the mid-infrared and P$\alpha$ images, the 2.2~$\mu$m
continuum image (Figure~2a) does not reveal a ring like structure.
Rather the 2.2~$\mu$m continuum shows a strong central peak, with an 
overall size similar to
the size of the mid-infrared image. This results from the fact that
the 2.2~$\mu$m light traces the stars which peak at the nucleus, while the
mid-infrared,  radio and P$\alpha$ images trace the ring of current
star fomation.

Figure~2c shows the SED of NGC 1614, comparing the photometric
measurements obtained with the Keck images in 2$''$ and  4$''$ diameter
beams, and the IRAS observations.  This plot shows that the Keck
observations detect $\sim$87\% of the flux  measured at 12~$\mu$m with
IRAS in a 4$''$ beam, and 72\% of the IRAS 12~$\mu$m flux in a 2$''$
beam.  The photometry shows that the mid-infrared emission is
predominantly confined to a nuclear region $\sim$ 1.7$''$ (550 pc) in
diameter (FWHM) with $<$30\% of the emission extending outside this
diameter.  The mid-infrared spectrum presented in Roche et al.  (1991)
shows a suggestion of both PAH emission and silicate absorption.  We do
not have adequate photometric data to distinguish PAH emission from
silicate absorption. The images at 11.7 and 12.5~$\mu$m show similar
structure, showing that grossly there are not significantly different
spectra in different locations in the starburst region.

\subsection {NGC 2623=Arp 243}

NGC 2623=Arp 243 is a strongly interacting galaxy at a redshift of
cz=5535 km/s (74 Mpc, 370pc/$''$) with an infrared luminosity of $3
\times 10^{11}L_{\odot}$ (Soifer et al. 1987). The optical image (Arp,
1966) shows two opposing tidal tails extending over 120$''$ (45 Kpc).
Near infrared NICMOS imaging (S00) shows a  single bright nucleus and
surrounding galaxy, and the 8.4~GHz  image (C91) shows a similar
structure  of  a bright compact core with an east-west elongated
structure.

The 12.5~$\mu$m image of NGC 2623 is shown in  Figure~3a, along with
the 8.4~GHz  and NICMOS 2.2~$\mu$m images. The angular resolution of
the 12.5~$\mu$m image is 0.7$''$, significantly lower resolution than
either the 8.4~GHz  (0.3$''$) or 2.2~$\mu$m (0.22$''$) images.  The
12.5~$\mu$m image has been registered with respect to the radio and
2.2~$\mu$m images by assuming the peak brightness at 12.5~$\mu$m agrees
with the peaks at the other wavelengths.

As can be seen from this figure, the structure of the 12.5~$\mu$m,
2.2~$\mu$m  and 8.4~GHz  images are all very similar and show similar
sizes. The observed FWHM size of the 12.5~$\mu$m source is 1.0$''
\times 0.7 ''$ or an intrinsic size of 0.7$'' \times < 0.35''$
(260$\times$$<$130 pc),  with the major axis oriented east-west. The
deconvolved size of the radio source is 0.43$'' \times
0.29''$(160$\times$110 pc). The overall extent of the 12.5~$\mu$m and
8.4~GHz emission agrees quite well, being $\sim 2 '' \times 1.5 ''$ (740
$\times$ 550 pc)  with the major axis oriented in the east-west
direction.  The excellent agreement between the 12.5~$\mu$m and
8.4~GHz  maps of NGC~2623 is illustrated directly in  Figure~3b, where
the 12.5~$\mu$m contours are overlaid on the grayscale image of the
8.4~GHz  image.  The 2.2~$\mu$m NICMOS  image shows a circular core
with a  size of 0.2$''$ (75 pc), and a larger nuclear region
having the same size and orientation as that seen at 12.5~$\mu$m and
8.4~GHz.

The flux densities presented in Table 2, and fluxes shown in
Figure~3c,  show that the 12.5~$\mu$m flux density measured in a 4$''$
diameter beam represents 80\% of the total flux density measured by
IRAS at 12~$\mu$m, while 65\% of the IRAS flux density is contained in
a 1$''$ diameter beam.  As in NGC 1614, the vast majority of the
mid-infrared emission in NGC 2623 is confined to the nuclear region
with a size of $<$400pc.  The 8--13~$\mu$m spectrum presented by
Dudley (1999) is consistent with the observations reported here, and
shows a PAH emission feature affecting the 11.7~$\mu$m flux
measurement. The mid-infrared spectrum suggests significant silicate
absorption of an  underlying PAH emission spectrum, though quantifying
this is highly  uncertain.

\subsection {NGC 3690+IC694=Arp 299=Mrk 171}

NGC 3690+IC 694 (Arp 299, Markarian 171) is one of the most
extensively studied interacting starburst galaxies known. Its optical
morphology (Arp 1966) shows two main bodies with many bright internal
knots and  diffuse tidal debris extending over 90$''$ (20Kpc). This
system was first shown to be a bright infrared and radio source by
Gehrz, Sramek and Weedman (1983) who showed 10~$\mu$m emission
extending over nearly 40$''$ (8 Kpc),  encompassing both galactic
nuclei. The redshift of NGC 3690 is cz=3120 Km/s (41 Mpc, 210
pc/$''$). IRAS measurements  showed a total  bolometric luminosity of
8$\times 10^{11}L_{\odot}$, making this the most luminous galaxy in
this study (Soifer et al. 1987), and  very close to the Ultraluminous
Galaxy limit (Sanders et al. 1988).

The optical spectrum is classified as coming from a HII region (e.g.,
Villeux et al. 1995) while the mid-infrared spectrum of Dudley (1999)
shows PAH emission and silicate absorption.  Sargent and Scoville
(1991) showed this system to be rich in molecular gas, while extensive
near infrared studies have been reported by Sugai et al.  (1999) and
Satyapal et al. (1999) among others. A detailed near infrared study of
NGC 3690 with NICMOS on HST has been reported by Alonso-Herrero et
al. (2000a). Previous groundbased mid-infrared imaging of NGC 3690 has
been reported by Miles et al. (1996) and Keto et al. (1997).

Figures 4a and 4b show the 12.5~$\mu$m image along with the NICMOS
2.2~$\mu$m, the 3.2~$\mu$m NIRC and the 8.4~GHz  images of NGC
3690/IC694.  The five nuclei identified at 2.2~$\mu$m are marked; the
marks are reproduced at the same relative positions in all four panels
of Figures 4a and 4b.  We use here the nomenclature for the  sources
introduced by Gehrz, Sramek and Weedman (1983), and Wynn-Williams et
al. (1991)  of A, B1, B2, C and C$'$.  Figure~4a shows the entire
extent of the emission, while Figure~4b shows detailed maps of the
western sources.  The astrometric registration of the maps at
different wavelengths is based on morphological similarities and
direct measurements. The 8.4~GHz  image of C91 and the 2.2~$\mu$m
image of Alonso-Herrero et al. (2000a)  were overlaid by assuming
spatial coincidence of sources A and B1.  The 2.2~$\mu$m and
3.2~$\mu$m images agree spatially to within 0.05$''$, based on NIRC
measurements.  After superposing the  12.5~$\mu$m image of source B1,
at 12.5~$\mu$m the western sources (B1, B2, C and  C') match the
locations at 3.2~$\mu$m to $<$0.05$''$.  Because of the extent of
NGC~3690, the nod amplitude was increased to 20$''$, and  all four
sources of NGC~3690W were thus imaged on one frame at 12.5~$\mu$m and thus
were registered accurately with respect to each other as seen in
Figure~4b. NGC~3690~A was imaged separately at 12.5~$\mu$m, with no other
objects in the field, and was registered with respect to the western
cluster in Figure 1a by assuming its peak emission coincides with the
2.2 $\mu$m and 8.4Ghz peaks.

The emission seen at 12.5~$\mu$m is localized into five distinct
emission centers at the locations of A, B1, B2, C and C$'$. 
The total flux density from the sum of the individual components is
reported in Table 2, and shows that the  total flux density
measured at 12~$\mu$m by IRAS is accounted for by the total of these
components as measured in the groundbased observations.  Thus at
least at 12~$\mu$m there is  negligible luminosity, in comparison to
these sources, being emitted in the many other supergiant HII regions
in this system.  A similar conclusion has been reached by Alonso-Herrero
et al. (2000a) based on the near infrared NICMOS observations of this
galaxy.

A direct comparion between the 12.5~$\mu$m and 8.4~GHz  maps of NGC
3690 is shown in  Figure~4c, where the contours of the 12.5~$\mu$m
image are overlaid on the grayscale of the 8.4~GHz  image of C91.
Sources B1, C and C' agree very well, while there is no apparent radio
counterpart for the faint 12.5~$\mu$m source B2.  The 12.5~$\mu$m
sources are quite compact; the Keck data have a resolution  of 0.6$''$
(125 pc), and only source C is clearly resolved in the Keck imaging.

While the vast majority of the infrared luminosity is produced by the
five sources isolated at 12.5~$\mu$m, there is a low level of thermal
infrared  emission that is detected clearly in the 3.2~$\mu$m image.
In  Figure~ 4d the 3.2~$\mu$m image is stretched to show the low
surface brightness emission, and for comparison the 1.644~$\mu$m
[FeII] image from Alonso-Herrero et al (2000a).  There is very low
level 3.2~$\mu$m emission seen  tracing the HII regions from source B2
to C, as well as the HII regions surrounding the nucleus in the
eastern galaxy (Source A = IC694).  

All of the sources detected at 12.5~$\mu$m are seen in the NICMOS
2.2~$\mu$m continuum images.  The source C$'$ is inconspicuous in the
near infrared image. This source has the largest ratio  of  $\frac
{S_{\nu}(12\mu~m)}{S_{\nu}(2.2\mu~m)}$ of those detected at
12.5~$\mu$m, being more than 5 times brighter at 12.5~$\mu$m  normalized
to the 2.2~$\mu$m flux density than either sources A or B1, the
brightest 12.5~$\mu$m sources.   Alonso-Herrero et al (2000a) do not
identify this as the source having the greatest extinction.  This
suggests that the more obscured sources (e.g. A, B1, C) have luminous
stellar contributions associated with but separate from the  highly
infrared luminous regions.

Figures 4e and 4f show the spectral energy distributions of the
various components. Figure~4e displays the SEDs of the integrated
light from the entire system, the sum of 4$''$ diameter beams centered
on A, B1, C and C', and a 2$''$ diameter beam centered on B2,  and the
individual SEDs in 2.5$''$ beams of the five separate peaks. Figure 4f
presents three panels that show the SEDs of sources A, B and C
separately in 1$''$ and 2.5$''$ diameter beams.  The 2.2$\mu$m data
are derived from NICMOS observations (Alonso-Herrero et al. (2000a),
while the 3.2$\mu$m data are from NIRC observations shown in Figure 4d.

As mentioned above, the total flux density measured
in the 12.5~$\mu$m images accounts for virtually all the flux
density measured by IRAS (Soifer et al. 1989). The spectroscopy
of Dudley (1999) shows that each of the separate sources, A, B1, C
and C$'$ includes a contribution from PAH emission, while sources
B1 and A also show overlying silicate absorption. The difficulty in
determining the contribution of PAH emission to the photometry is
illustrated by the comparison sources A and C in Figure~4d. Source A has
an apparent small PAH contribution, but the spectrum of Dudley
reveals that there is a substantial PAH emission feature superposed
on a steeply rising spectrum, while the similar strength feature
seen in the SED of source C, which is suggestive of PAH emission
in the photometry, is a result of being superimposed on a flatter SED for
this source.

\subsection {IC 883=UGC 8387=Arp 193}

IC883=UGC 8387=Arp 193 is  a disk galaxy with crossed tails in optical
images suggestive of a merger.  At a redshift of cz=7000 Km/s (92 Mpc,
470 pc/$''$), the infrared luminosity of this system is 3$\times
10^{11}L_{\odot}$ (Soifer et al. 1987). Its optical spectrum (Villeux,
et al. 1995) is classified as that of a LINER, while its mid-infrared spectrum
shows PAH emission (Dudley 1999).  High resolution NICMOS observations
of the bright central portion of this system reveal an apparently edge
on disk galaxy with a significant dust lane (S00).  High resolution
radio observations show a source extended along the infrared disk
with a size of $\sim 4'' \times0.7 ''$ (C91).

 Figure~5a shows the 12.5~$\mu$m Keck image, along with the 2.2~$\mu$m
NICMOS  image from S00 and the 8.4~GHz  radio image from C91.  The
12.5~$\mu$m image of  Figure~5a is smoothed to 0.62$''$ resolution from
the original 0.40$''$ to improve the signal-to-noise ratio, while the
resolution is 0.19$''$ at 2.2~$\mu$m and 0.24$''$ at 8.4~GHz.  Two peaks
visible at 2.2~$\mu$m are marked; the marks are reproduced at the same 
relative positions in all three panels of Figure~5a.

The astrometric registration of the radio and 12.5~$\mu$m images is
based on the assumption that the 12.5~$\mu$m peak coincides with the
southeastern  radio peak. This forces general agreement between the
radio and 12.5~$\mu$m structures. The 2.2~$\mu$m image was registered
with respect to the radio image by assuming coincidence of the central
peak at 2.2~$\mu$m and the southeastern radio peak.

The 12.5~$\mu$m image shows that the emission is extended along a
position angle of  135~\degr and shows a significant extension
perpendicular to the major axis.  The measured 12.5~$\mu$m FWHM is
1.8$'' \times 0.7''$ while the full extent of the 12.5~$\mu$m emission
is 3$'' \times 1.7''$.  The 8.4~GHz  image shows a similar size along
the major axis with significantly less extent perpendicular to this
axis, having a full extent along the  minor axis of 0.7$''$.  With the
assumption that the southeastern radio peak coincides with the peak of
the 12.5~$\mu$m emission,  the northwest radio peak appears to coincide
with a shoulder of 12.5~$\mu$m emission. This is illustrated in
Figure~5b where the 12.5~$\mu$m contours are overlaid on a grayscale of
the 8.4~GHz  image.  The emission perpendicular to the major axis
appears significantly more extended at 12.5~$\mu$m than at 8.4~GHz ,
but this could be a result of the lack of sensitivity to low
surface brightness emission in the radio image.

The 2.2~$\mu$m image does not agree in morphology particularly well
with the mid-infrared and radio images.  Again we have assumed that
the central peak of 2.2~$\mu$m emission coincides with the radio and
infrared peaks, and present in  Figure~5b a grayscale image of the
NICMOS 2.2~$\mu$m image with 12.5~$\mu$m contours overlaid.  The
2.2~$\mu$m peak to the northwest of the center differs in position
relative to the radio peak by 0.2$''$ or 90 pc at the distance of the
galaxy. In addition, the 2.2~$\mu$m extent is significant to the
southeast of the central peak, while both the mid-infrared and 8.4~GHz
images show extent to the northwest of the central peak.

In addition to imaging IC~883 at 12.5~$\mu$m, Keck images were
obtained at 8.0, 10.3 and 11.7~$\mu$m. These images showed virtually
identical morphology to that shown in  Figure~5a. Based on  one
dimensional  profiles of flux vs.  position from each Keck image there
is no discernable variation in  the SED of IC~883 along its major axis
at a resolution of 0.6$''$ (270 pc)  over the central 3$''$ or 1.4
Kpc.  The spectral energy distribution of IC~883 is shown in Figure
5c.  The comparison of the Keck observations and the IRAS data shows
that the Keck observations account for all of the mid-infrared
emission detected by IRAS in this system. The comparison of the 1$''$
and 4$''$  photometry illustrates the point that the SED does not vary
significantly with  beam diameter.

The Keck data show a strong drop in flux at 10~$\mu$m, indicative
of significant silicate absorption affecting the emergent
spectrum. The mid-infrared spectrum of Dudley (1999)  shows strong PAH
emission with a depression that is not nearly as strong as presented
here. This apparent discrepancy is likely due to the low
signal-to-noise ratio in the Dudley observations at $\sim$10~$\mu$m.

\subsection {NGC 6090 = UGC 10267}

NGC 6090 = UCG 10267, at a redshift cz = 8785 km/s (117 Mpc, 590
pc/$''$), has an infrared luminosity of 3$\times 10^{11}L_{\odot}$
(Soifer et al. 1987).  The visual image shows a face-on spiral with a
close companion separated by 5$''$ or $\sim$3 Kpc to the southwest and
faint tidal tails extended over $\sim 2'$ (72 Kpc).  The  optical
spectrum is classified as HII like (Veilleux et al. 1995).  High
resolution VLA imaging at 8.4~GHz  (C91) shows a compact source, while
a lower resolution map at 1.4~GHz  (Condon et al. 1990) shows emission
extending over the face of both the spiral galaxy and the close
companion.  NICMOS observations have been reported for the NGC 6090
system by S00 and Dinshaw et al. (2000).  Bryant and Scoville (1999)
found that the molecular gas in this system is located approximately
midway between the two galaxies.

The 12.5~$\mu$m image of NGC~6090 is presented in  Figure~6a, along
with the 2.2~$\mu$m image from S00, an 8.4~GHz  image reprocessed from
the data of C91 with angular resolution of 0.5$''$   and the
Br$\gamma$ image from Soifer et al. (2001). The astrometric
registration of the radio and 12.5~$\mu$m images is based on the
assumption that the brightest 12.5~$\mu$m peaks coincide with radio
peaks with the same angular separation (and position angle).  The
mid-infrared image was located with respect to the near infrared
continuum through the Br$\gamma$ image.  Because of the strong
similarity in the morphological structure between the 12.5~$\mu$m and
Br$\gamma$ images, the 12.5~$\mu$m emission peaks  were assumed to
coincide with corresponding peaks in the Br$\gamma$ image. The
Br$\gamma$ image was obtained using the Palomar Integral field
Spectrograph (PIFS, Murphy et  al. 1999) and was obtained
simultaneously with a 2.2~$\mu$m continuum image of this galaxy. The
2.2~$\mu$m continuum image obtained with PIFS corresponds well with
the 2.2~$\mu$m NICMOS image, so that the Br$\gamma$ and 2.2~$\mu$m
continuum images are located very precisely ($< 0.08''$) with respect
to each other.  The PIFS data will be  discussed in more detail in
Soifer et al.

In addition to the morphological similarities between the 12.5~$\mu$m
and Br$\gamma$ images there is a strong physical reason for expecting
the emission to be  spatially coincident; the currently ionzed gas
traces the ionizing stars and the mid-infrared emission traces the
dust heated by these (presumably) same stars.

With the {\it assumption} that the 12.5~$\mu$m and Br$\gamma$ images
coincide, a faint 12.5~$\mu$m peak to the east of the brightest
regions in the image, coincides with the 2.2~$\mu$m nucleus in the
face-on spiral NGC 6090. This is shown directly in Figure~6b, where
the 12.5~$\mu$m contours are overlaid on a grayscale of the 2.2~$\mu$m
image.  The main emission at 12.5~$\mu$m coincides with regions of
blue, unresolved sources in the spiral arms in the face-on spiral
galaxy as noted by S00. In addition, the faint
12.5~$\mu$m emission $\sim$3$''$ south and east of the bright
12.5~$\mu$m emission is located on a minor peak (not the bright point
source) in the nearby companion galaxy. This region is  spatially
coincident with additional blue unresolved sources in the companion
galaxy  (S00).  The bright unresolved source in the companion galaxy
does not coincide with any radio or mid-infrared emission, consistent
with the suggestion that it is a foreground object.  The brightest
mid-infrared peaks are extended by $\sim 1''$, or  600 pc, while the
whole 12.5~$\mu$m emitting region in NGC 6090 is extended over 4$''
\times 2''$, or 2.4 $\times$ 1.2 kpc.

With the astrometric registrations as presented in Figure 6a, the
agreement between the peaks of 8.4 GHz emission and 12.5$\mu$m
emission is quite good.  Each peak of 12.5$\mu$m emission corresponds
to a peak of  radio emission (with perhaps a 0.2$''$ discrepancy
between the faint 12.5$\mu$m peak to the southeast and the
corresponding radio peak). In addition, there appears to be a radio
peak at the position of the near infrared nucleus of NGC 6090.

In NGC~6090 the galaxy nucleus is at best a minor source of infrared
luminosity. The current location of luminous star formation appears to
be in the spiral arms of the face-on galaxy and the portions of the
companion galaxy closest to the spiral. These regions are within
spiral arms  and close to the molecular gas which lies between the two
galaxies (Bryant and Scoville, 1999).  This suggests that the
molecular gas  remained behind after the passage of the two galaxies
on a closely interacting trajectory, and the current star formation
was  triggered in the locations of closest approach in the two
galaxies.

Figure~6c shows the spectral energy distributions of the peaks in
NGC~6090, along with the large beam IRAS and 2MASS data.  At 12~$\mu$m
approximately 50\% of the flux density measured by IRAS is detected in
the LWS images.  The substantial difference between the 12~$\mu$m flux
density measured by IRAS and the integrated flux density measured at
12.5~$\mu$m in the Keck image argues that there is significant diffuse
mid-infrared emission in this system not detected in the ground-based
imaging. Keck observations were obtained  at 11.7, 12.5 and
17.9~$\mu$m. The increased flux density at 11.7~$\mu$m compared to
that at 12.5~$\mu$m as seen in Figure~6c suggests that there is strong
PAH emission in this system. There is no published spectrum of NGC
6090 spanning this wavelength range.

\subsection {Markarian 331= UGC 12812}

Markarian 331 = UCG 12812, at a redshift cz = 5500 km/s (73 Mpc, 370
pc/$''$), has an infrared luminosity of 2.5$\times 10^{11}L_{\odot}$
(Soifer et al. 1987).  The visual image shows a high surface
brightness core with no obvious tidal tails.  There are two galaxies
of comparable brightness located within 2$'$.  The
optical spectrum is classified as HII like (Veilleux et al. 1995),
with a mid-infrared  spectrum indicating a combination of PAH emission
and silicate absorption (Dudley, 1999).  High  resolution VLA imaging
at 8.4~GHz  (C91) shows a  bright central source  surrounded by an
elliptical ring of emission with dimensions 3$'' \times$2$''$.

Figure~7a shows the 12.5~$\mu$m image of Markarian 331 along with
images at 2.15~$\mu$m (K$_s$), 3.4~$\mu$m and 8.4~GHz.  A
bright nucleus is seen at all wavelengths,  as well as extended
emission in an apparent disk or ring.  In registering the images, it
was assumed that the central peak was spatially coincident at all
wavelengths.

At 8.4~GHz, the extended emission is  distributed as a ``ring"
surrounding the bright nucleus. At 12.5~$\mu$m there is a bright
nucleus  that is unresolved in the 0.3$''$ seeing, as well as emission
that appears to have a  similar structure as the radio ring.  The
12.5~$\mu$m emission appears to be distributed over the full disk  of
the system, extending over $\sim4''$ or 1.5 Kpc diameter, with a
``bar-like" structure extending from the nucleus to the northeast. The
image of Markarian 331 obtained at 11.7~$\mu$m is virtually identical
to that at 12.5~$\mu$m. At 8.4~GHz  the emission is apparently confined
to the nucleus and the ring. The  latter effect may be a result of the
very high resolution of the VLA image.  The structure at 3.4~$\mu$m
appears quite similar to that at 12.5~$\mu$m, showing a compact
nucleus, a bar to the northeast and a disk.   Figure~7b compares the
images in the mid-infrared and radio directly, with an overlay of the
12.5~$\mu$m contours on a grayscale 8.4~GHz  image.  The structures are
quite similar, with the peaks in the radio image corresponding very
closely to the features in the mid-infrared image.

Figure~7c plots the fluxes in this system, comparing the
mid-infrared flux densities to the IRAS and 2MASS measurements.  The
ground-based data  account for $>$80\% of the 12~$\mu$m flux density
measured by IRAS, showing that the central $\sim$Kpc accounts for
nearly all the infrared luminosity in this system.  The similarity of
the images at 11.7 and 12.5~$\mu$m shows that there is little evidence
for variation in the spectral distribution of the infrared emission
over the nuclear region. Roche et al. (1991) show that there is a
strong PAH feature in the 10~$\mu$m spectrum of this galaxy, as well as
strong silicate absorption.

\section {Discussion}

The galaxies observed in this study were selected to be highly
luminous, active star forming galaxies that are  bright in the
12~$\mu$m IRAS band.  The mid-infrared observations reported here
generally account for well more than half of the total 12~$\mu$m flux
density as measured by IRAS.  This is seen in the figures of SEDs of
the observed galaxies, where the observed mid-infrared fluxes are plotted
with the IRAS fluxes. Only in NGC 6090 and VV114 is the
12.5~$\mu$m flux observed here less than half the IRAS
12~$\mu$m flux.  In the remaining five systems, the the
compact structures observed in the Keck images account for 80--100\%
of the total 12~$\mu$m flux density from the system.  Thus our first
conclusion is that star formation in these systems is not occuring
over the entire disk of the galaxy, but is confined to relatively
compact regions having sizes of $\sim$100 pc  up to $\sim$ 1 Kpc.

The mid-infrared emission in these systems, as characterized by the
12.5~$\mu$m images presented in Figures 1--7, is far more compact
than the overall distribution of  near infrared light in these
systems. This can be  seen by comparing the 12.5~$\mu$m images of
Figures 1a -- 7a  with the corresponding large field 2.2~$\mu$m NICMOS
images from S00.  The sizes of the mid-infrared sources are typically
$\sim$ 0.3$''$ --2$''$, while the galaxy sizes seen in the NICMOS
images (S00) are typically 10$''$--30$''$.  The significantly smaller
mid-infrared sizes compared to the near infrared sizes of these
sources  is  shown directly in Figure 8, where we present the
curves-of-growth of the 12.5~$\mu$m flux as a function of beam radius,
compared to those at 2.2~$\mu$m and 1.1 (or 1.2)~$\mu$m for the
galaxies in the sample. The 12.5~$\mu$m flux vs. beam size was
determined from the images in Figures 1a -- 7a, with  the
normalization to 100\% provided by the IRAS measured total flux
density at 12 $\mu$m. The origin for the measurement beams was the
location of peak 12.5~$\mu$m brightness as identified in the galaxy
images of Figures 1a -- 7a.

The corresponding curves-of-growth at 1.1 and 2.2~$\mu$m were
determined from  NICMOS images for VV114, NGC 2623, IC 883, NGC 6090
(from S00) NGC 1614 (from  Alonso-Herrero et al. 2000b) and NGC 3690/
IC 694 (from Alonso-Herrero et al. 2000a) and from Palomar J and K$_s$
imaging data for Markarian~331. The large beam normalizations for
these data were based on the total J and K magnitudes of these
galaxies as measured in the 2MASS database (T. Jarrett, private
communication). For the 1.1 and 2.2~$\mu$m curves-of-growth, the
origin for the measurement beams was the same as for the 12.5~$\mu$m
beams, as shown in Figures 1a -- 7a.

Figure 8 shows that for the galaxies where the ground-based
12.5~$\mu$m images account for the bulk of the total 12~$\mu$m flux of
the galaxy, the radii of the regions producing 50\% of the flux at
12.5~$\mu$m  are $<$100~pc to $\sim$300~pc, while the equivalent radii
at 1.2 and 2.2 $\mu$m  are 1 to $>$2 Kpc.  Noticeably different from this
general characteristic are VV114 and NGC 6090, where a substantial
fraction ($\sim$60\%)of the 12~$\mu$m light is not detected in the
ground-based observation, and is presumably extended, low surface
brightness  emission distributed over the galaxies.  For both VV114E
and  NGC 6090 the curves-of-growth at 2.2 and 12.5~$\mu$m are quite
similar.

The 2.2~$\mu$m light distribution  is, in principle, relatively
insensitive to dust extinction, although this is not the case for very
dusty environments in the galaxies. To the  extent that older stellar
populations dominate the near infrared light in these galaxies, the
2.2~$\mu$m light traces the mass distribution of that population.

The 12~$\mu$m emission traces the current sites of luminous star
formation.  If the star forming regions within these galaxies are
local examples of  distant, dusty star forming galaxies, the global
SEDs of the distant star-forming galaxies would be more like those of
the starburst regions within these galaxies, rather than like the SEDs
of the whole galaxies, which have major contributions in the near
infrared from stars not associated with the current starburst.  The
difference between the global SEDs and the SEDs of the star forming
regions is illustrated quantitatively in the figures  by the
comparison of the integrated SEDs with the small aperture SEDs of the
individual 12.5~$\mu$m sources in the galaxies.  In all the galaxies,
the SED of the individual star forming regions shows a significant
increase in ratio of 12.5~$\mu$m to 2.2~$\mu$m flux in the small
beams, compared to the integrated light of the galaxies. Typically
this ratio is larger by a factor of two  in the star forming region as
compared to the integrated light for the galaxy, but it ranges to
nearly an an order of magnitude for the reddest source (C$'$) in NGC
3690.  For distant star forming galaxies, such as those seen in SCUBA
submillimeter surveys (e.g. Eales et al. 2000), SIRTF observations
will be required to obtain the equivalent observations to those
presented here.

Presumably the 12.5~$\mu$m emission is tracing the far infrared
emission associated with the same star formation complexes.  If the
fraction of the total infrared luminosity emitted by the individual
regions in these galaxies is similar to the fraction of the
12.5~$\mu$m flux density compared to the total 12~$\mu$m flux density
measured by IRAS, these regions  produce from $\sim
10^{10}$L$_{\odot}$ to $\sim 4\times 10^{11}$L$_{\odot}$ within a few
hundred pc.  While not as luminous as the ULIRGs, the star formation
rates are still  prodigious by comparison to normal galaxies like the
Milky Way. The star formation rates are from 0.7 to  30
M$_{\odot}$/yr,  based on the conservative luminosity/star formation
rate conversion of Scoville and Young (1983) or Scoville and Soifer
(1990).  If the conversion from luminosity to star formation rate is a
factor of several greater than this, as is the case for models that
account for the formation of stars with mass $<$1 M$_{\odot}$(cf.
Kennicutt, 1998), the total star formation rates go up accordingly.

Since the mid-infrared sources trace the high luminosity regions in
these galaxies, these data provide direct evidence of the physical
characteristics of the regions in which the starburst luminosity is
being generated.  Table 3 presents a summary of the sizes, brightness
temperatures, and color temperatures of the mid-infrared sources. The
inferred brightness temperature at 60~$\mu$m is calculated assuming
the same angular size as measured at 12.5~$\mu$m,  and assuming the
same fraction of the total 60~$\mu$m flux density (as determined from
the IRAS data) emerges from the region as is observed for the source
at 12~$\mu$m.  In physical sizes, the detected sources range from $<
110$ pc to almost 1 Kpc.  In a few cases it is not physically possible
for the bulk of the luminosity to emerge from sources of the sizes
inferred from the 12.5~$\mu$m images, since the brightness temperature
at 60~$\mu$m is larger than the color temperature of the source (here
we assume that the color temperature is the global color temperature,
determined from the IRAS 60 and 100 $\mu$m flux densities). In these
cases the far infrared source size must be larger than the observed
12.5~$\mu$m source size.  The minimum source sizes necesary for equal
color and brightness temperatures at 60~$\mu$m are also given in Table
3, and generally are close to, or slightly larger than, the upper
limits on sources sizes determined from the 12.5~$\mu$m Keck
observations.

With these estimates of the physical sizes of the emitting regions
determined, we can compare the derived surface brightnesses with those
of both lower and higher luminosity sources.  Table 3 includes the
surface brightnessess in L$_{\odot}$/Kpc$^2$ derived from the FWHM
estimated source sizes (corrected for the PSF size) and the fraction
of the total luminosity of the source inferred from the 12.5~$\mu$m
measured flux density for that source.  As can be seen from Table 3,
these values range from 2$\times 10^{11}$ to $\sim 2\times 10^{13}
$L$_{\odot}$/Kpc$^2$.  By comparison, the infrared surface
brightnesses in the ULIRGs observed by Soifer et al. (2000) range from
1$\times 10^{12}$ to $\sim 6 \times 10^{13}$ L$_{\odot}$/Kpc$^2$.  In
physical size, the starburst regions range from $\sim 120$ pc (in
VV114E$_{NE}$) to $\sim$ 1.2 Kpc (in Markarian 331), while the
corresponding sizes of starburst regions in ULIRGs are quite similar,
ranging from 140 pc in Arp 220 to 1.6 Kpc in IRAS 17208-0014. Thus,
while the apparent physical sizes are similar to those found in
ULIRGs, the apparent surface brightnesses of star formation are lower
by factors of 3--10 or more.  The  difference in luminosities between
the luminous starburst galaxies and the ULIRGS appears to be in the
luminosity generated in a given volume rather than in the volume in
which  the luminosity is generated.

The core of Orion, with a luminosity of $2 \times 10^5$ L$_{\odot}$
over 0.3 pc (Werner et al. 1976), has an apparent surface brightness
of $3\times 10^{12}$ L$_{\odot}$/Kpc$^2$, while the apparent face-on
surface brightness of the starburst region in M82 is $\sim 2\times
10^{11}$L$_{\odot}$/ Kpc$^2$ (Soifer et al. 2000).   Previous
work (Meuer et al. 1997) has shown that in normal galaxies, the
surface brightness of star formation  reaches an upper limit of $\sim
2 \times 10^{11}$L$_{\odot}$/Kpc$^2$ globally averaged over entire
galaxies, with peak surface brightnesses of clusters of size $\sim$10
pc of $\sim 5 \times 10^{13}$L$_{\odot}$/Kpc$^2$.  Thus the starburst
regions in these systems appear extraordinary in comparison with
normal galaxies, but not with respect to the highest density
star forming environments within galaxies.

Table 4 summarizes the surface brightnesses of nearby starburst systems,
the starburst regions in the galaxies in this sample, and the starburst
regions in ULIRGs from Soifer et al. (2000).  We can see clearly here
that while there is significant scatter, there appears to be a natural
progression in surface brightness in the galaxies, with increasing surface
brightness accompanying increasing luminosity. Table 4 shows that in these
luminous starburst galaxies environments equivalent to the core of Orion 
extend over hundreds of parsecs. 

Turner, Beck \& Ho (2000) and Gorjian, Turner \& Beck (2001)  have
recently shown that in the nearby dwarf galaxy NGC 5253 star formation
that produces $\sim$ 2$\times 10^9$L$_{\odot}$ is occurring  in a
region $\sim$ 2 pc in size, i.e. in a ``super starcluster''. The
observations described here cannot rule out the possibility  that much
of the luminosity in the starburst galaxies studied here emerges from
regions of similar size, although the most luminous systems in this
sample would require clusters of $\sim$100 such ``super starclusters''
in regions of order $\sim$200 pc in diameter.

In about half of the galaxies in this sample the luminosity appears to
be  generated in nuclear starbursts. In NGC 1614 and Markarian 331, a
substantial fraction of the infrared luminosity is generated in a ring
surrounding the nucleus.  NGC 6090 presents the greatest exception to
the nuclear starburst picture for high luminosity systems. Here the
luminosity appears to be generated where the molecular gas remains
from the recent interaction  between the two galaxies.  This might be
the best example for a starburst triggered by cloud-cloud collisions
in the galaxies disks.

The mid-infrared and radio emission generally trace each other quite
well.  Differences at the 100 pc scale are seen in VV114E$_{SW}$ and in
the more extended  emission in  NGC 6090 and IC 883.  It is most
likely that the differences in the diffuse emission is a lack of
sensitivity to low surface brightness emission in the 8.4 Ghz data.

\section {Summary and Conclusions}

We have obtained mid-infrared diffraction limited imaging of seven
starburst galaxies with the  Keck Telescopes. These observations have
shown

1 - A substantial fraction, usually more than 50\%,  of the mid-infrared
luminosity in these systems is generated in regions ranging in size
from $\sim$100 pc to 1 Kpc,

2 - Nuclear starbursts generally  dominate the starburst activity in these
galaxies.

3 - In some cases, most notably NGC 6090, substantial infrared
luminosity is generated in regions away from the galaxy nuclei, and is
more likely associated with the region of physical interaction between
two galaxies.

4 - The radio emission in starburst galaxies is a good tracer of the
current location of compact star-formationing regions.

5 - Mid-infared emission is much more compact than is near infrared 
emission in these galaxies. If the starburst regions in these galaxies
are prototypes for starforming galaxies at high redshift, the energy output
of such galaxies should be dominated by the mid and far-infrared energy
output more than is the case in the integrated light emerging from
nearby starburst galaxies.

\section {Acknowledgments}

We thank  J. Aycock, T. Stickel, G. Wirth, R. Campbell and Lee Armus
for assistance with the observations.  Barbara Jones, Rick Puetter and
the  Keck team brought the Long Wavelength Spectrograph (LWS)  into
service, enabling many of these observations. We thank Marcia Rieke
and Susan Neff for providing data for NGC 3690, NGC 6090 and NGC
1614. Vassilis Charmandaris provided advice about ISO observations of
VV 114. Joe Mazzarella provided us with data from NED, Tom Jarrett
provided us with data from 2MASS, and Aaron Evans provided images from
NICMOS. We also thank an anonymous referee for critical comments on
this paper.

The W. M. Keck  Observatory is operated as a scientific partnership
between the California Institute of Technology, the University of
California and the National Aeronautics and Space Administration. It
was made possible by the generous financial support of the W. M. Keck
Foundation.  This research has made use of the NASA/IPAC Extragalactic
Database which is operated by the Jet Propulsion Laboratory, Caltech
under contract with NASA.

B.T.S, G.N., K.M. and E.E. are  supported by grants from the NSF and
NASA.  B.T.S. and S.S are supported by the SIRTF Science Center at
Caltech.  SIRTF is carried out at the Jet Propulsion Laboratory.
N.Z.S. was supported by NASA grant NAG5-3042.
This work was carried out in part (M.R.) at the Jet Propulsion
Laboratory, operated by the California Institute of Technology, under
an agreement with NASA. The development of MIRLIN was suported by
NASA's Office of  Space Science.  The National Radio Astronomy
Observatory (J.J.C.) is a facility of the National Science Foundation
operated under cooperative agreement by Associated Universities, Inc.
This research has made use of the NASA/IPAC Extragalactic Database (NED),
which is operated by the Jet Propulsion Laboratory, California Institute 
of Technology, under contract with NASA.

\newpage

\figcaption{}  Figure~1a A montage of contour plots is presented of the
VV114E at 2.2~$\mu$m, 3.2~$\mu$m, 12.5~$\mu$m and 8.4~GHz. In this, and
all subsequent figures, north is up and east to the left, and the FWHM
size of the point spread function is shown separately in the panel for
each wavelength as the hatched circle. In  this, and all subsequent
figures, contour levels are spaced by a factor of 1.34. The first
contour is set to 90\% of the peak brightness so that the third contour
down represents 50\% of the peak brightness. In  this, and all
subsequent figures, unless otherwise indicated the 2.2~$/mu$m images
are from S00, while the radio images are from C91. In this, and all
subsequent figures that display contour plots of the galaxies, the
horizontal bar in one panel represents 500 pc at the galaxy.  In
Figure~1a the origin is chosen to be at the location of the northeast
peak in the 2.2~$\mu$m image. The astrometric registration of the
images at the different wavelengths is discussed in the text. The
locations of the brightest peaks in the 2.2~$\mu$m image are indicated
by a + and $\times$ in the same location in each panel.

 Figure~1b  A contour map of VV114E at 12.5~$\mu$m overlaid on the
corresponding grayscale image of 8.4~GHz  emission in the left panel.
An overlay of the 3.2~$\mu$m emission contours on the corresponding
grayscale image of 8.4~GHz  emission is  presented in the center
panel.  An overlay of the 12.5~$\mu$m emission contours on the
corresponding grayscale image of the 2.2~$\mu$m emission is  presented
in the right panel. In these, and all subsequent gray scale images,
the wavelength of the contoured image  is given in the lower left
corner and that of the gray scale image in the lower right.

 Figure~1c A contour map of the full VV114 field at 3.2~$\mu$m
overlaid  on a gray scale image of the same region at 2.2~$\mu$m.

 Figure~1d  The spectral energy distributions for (top panel) all of
VV114, VV114E, VV114E$_{NE}$ and  VV114E$_{SW}$  and (bottom panel)
all of VV114 repeated and  VV114W. The data are plotted as flux per
octave, $\nu${\it f}$_{\nu}$, vs. wavelength.  In the top panel the
integrated fluxes, taken from the IRAS and 2MASS data are shown as
filled circles, along  with photometry in a a 4$''$ diameter beam
centered on VV114E (filled squares). In addition,  photometry in 1$''$
diameter beams centered on VV114E$_{SW}$ ($\times$) and
VV114E$_{NE}$(+) as indicated in figure 1a is also shown as open
squares.  In this, and all subsequent SEDs, the beam diameter for the
photometry is indicated next to the identifying symbol. In this and
the subsequent SEDs, the photometric points made with the same beam
size are joined together with straight lines intended only to guide
the eye. They are not intended to indicate the absence of significant
structure between the data points presented, e.g., due to PAH emission
or silicate absorption.  The mid-infrared  photometry is from data
presented here. The 2.2~$\mu$m data are from the NICMOS imaging of
S00, while the 3.2~$\mu$m data are from NIRC.  In the bottom panel the
integrated fluxes are repeated,  as well as the measured emission for
VV114W in a 4$''$ diameter beam centered on the western peak shown in
Figure~1c.

\figcaption{}  Figure~2a A montage of contour plots of NGC 1614 at
P$\alpha$(1.87~$\mu$m), 2.2~$\mu$m, 12.5~$\mu$m, and 4.8~GHz is
presented.  The  origin is chosen to be at the peak emission in the
2.2~$\mu$m image.   The astrometric registration of the images at the
different wavelengths is discussed in the text. The location of the
center in each image is indicated by a + in each panel. The 2.2~$\mu$m
broadband and the 1.87~$\mu$m P$\alpha$ NICMOS images are from
Alonso-Herrero et al. (2000b).

 Figure~2b  A contour map of NGC 1614 at 12.5~$\mu$m is overlaid on
the corresponding grayscale image of P$\alpha$ emission in the left
panel.  An overlay of the 12.5~$\mu$m emission contours on the
corresponding grayscale image of 4.8~GHz  emission is  presented in
the right panel.

 Figure~2c  The spectral energy distribution of NGC 1614 is shown.  The
filled circles represent the integrated fluxes taken from IRAS and
2MASS data.  The aperture photometry in a 4$''$ diameter beam is
represented by filled squares, the photometry in a 2$''$ diamater beam
is represented by open circles. The 2$''$ diameter was set to include
the ring of emission seen in Figure 2b. The 2.2~$\mu$m photometric data
are from the NICMOS image on NGC 1614 (Alonso-Herrero et al., 2000b) and
Carico et al. (1988).

\figcaption{}  Figure~3a A montage of contour plots is presented of
NGC 2623 at 2.2~$\mu$m, 12.5~$\mu$m, and 8.4~GHz. The origin is chosen
to be at  the peak emission in the 2.2~$\mu$m image.  The astrometric
registration of the images at the different wavelengths is discussed
in the text. The location of the center in each image is indicated by
a + in each panel.

 Figure~3b  A contour map of NGC 2623 at 12.5~$\mu$m is overlaid on the
corresponding grayscale image of 8.4~GHz  emission.

 Figure~3c  The spectral energy distribution of NGC 2623 is shown.  The
filled circles represent the integrated fluxes taken from IRAS and
2MASS data.  The aperture photometry in a 4$''$ diameter beam is
represented by filled squares, the photometry in a 1$''$ diamater beam
is represented by open circles.  The 2.2$\mu$m photometric data are
scaled from  (S00).

\figcaption{}  Figure~4a A montage of contour plots is presented of NGC
3690/IC 694 at 2.2~$\mu$m, 3.2~$\mu$m, 12.5~$\mu$m and 8.4~GHz. The
positions of  sources A, B1, B2, C and C$'$ as determined at 2.2~$\mu$m
are indicated in each panel.  The astrometric registration of the
images at the different wavelengths is discussed in the text. The
2.2~$\mu$m image is from Alonso-Herrero et al. (2000a) and the
3.2~$\mu$m image is from NIRC.

 Figure~4b A montage of contour plots of NGC 3690 at  2.2~$\mu$m,
3.2~$\mu$m, 12.5~$\mu$m and 8.4~GHz is given. This is an expanded
scale of the western portion  of Figure~4a, showing in more detail the
sources in NGC 3690, but otherwise the same as in  Figure~4a.

 Figure~4c  A contour map of NGC 3690 at 12.5~$\mu$m overlaid on the
corresponding grayscale image of 8.4~GHz  emission is shown.  

 Figure~4d Grayscale images of NGC 3690/IC 694 at 3.2~$\mu$m and in the
[FeII] 1.644~$\mu$m emission line + continuum (from Alonso-Herrero et
al. 2000a) are shown.  Both images has been stretched to enhance the
low level emission.

 Figure~4e The spectral energy distributions of sources in NGC 3690 /
IC 694 are presented.  The integrated fluxes are represented by filled
circles, and are taken from IRAS and 2MASS data.  The filled squares
represent  photometry in a 2.5$''$ diameter beam centered on sources
A, B1, C and C$'$ and a 2$''$ diameter beam centered on source B2 as
shown in Figures 4a and 4b. The filled triangles represent the sum of
the photometric points for the individual sources using a 4$''$ beam 
for sources A, B1, C, and C$'$ and a 2$''$ beam for B2.  The 3.2~$\mu$m
data  are from NIRC, while the 2.2~$\mu$m data for the small
beams  are from NICMOS (Alonso-Herrero et al. 2000a).

Figure~4f The spectral energy distributions of individual sources in
NGC 3690 / IC 694.  The filled triangles represent the
sum of all the photometric points for the individual sources, and are
presented for reference in each panel. In the top panel are data for
source A in 1$''$ (filled squares) and 2.5$''$ (open squares) diameter
beams.  The middle panel presents data for sources  B1 (1$''$diameter
beam - open squares, 2.5$''$diameter beam - filled squares) and B2
(1$''$diameter beam - open squares, 2$''$diameter beam - filled
stars).  The bottom panel presents data for sources C (1$''$diameter
beam -  open squares, 2.5$''$diameter beam - filled squares) and C$'$
(1$''$diameter  beam - open squares, 2.5$''$diameter beam - filled
squares).  The data sources are as in Figure~4e.

\figcaption{}  Figure~5a A montage of contour plots is presented  of
IC 883 at 2.2~$\mu$m, 12.5~$\mu$m and 8.4~GHz.  The origin is chosen
to be at the  peak emission in the 2.2~$\mu$m image.  The astrometric
registration of the images at the different wavelengths is discussed
in the text. The location of the two brightest peaks at 2.2~$\mu$m are
indicated in each image by a + and $*$ at the same relative location
in each panel.

 Figure~5b An overlay of the 12.5~$\mu$m emission contours of IC 883
on the corresponding grayscale image at 2.2~$\mu$m is presented in the
left panel. A contour map  at 12.5~$\mu$m is overlaid on the
corresponding grayscale image of 8.4~GHz  emission in the right panel.

 Figure~5c  The spectral energy distribution of IC 883 is shown.
The filled circles represent the integrated fluxes taken from IRAS and
2MASS data.  The aperture photometry in a 4$''$ diameter beam is
represented by filled squares, the photometry in a 1$''$ diamater beam
is represented by open circles.  The 2.2~$\mu$m photometric data are 
scaled from the NICMOS image (S00).

\figcaption{}  Figure~6a A montage of contour plots is presented of
NGC 6090 at Br$\gamma$ (2.16~$\mu$m), 2.2~$\mu$m, 12.5~$\mu$m, and
8.4~GHz.  The origin is chosen to be at the nucleus of the face-on
spiral galaxy in the 2.2~$\mu$m image, and is indicated by a $+$ in
each panel.  The astrometric registration of the images at the
different wavelengths is discussed in the text. The location of a
secondary peak at 2.2~$\mu$m is  indicated by a $*$ in each panel.
The location of the secondary peak in  the 8.4 GHz emission is
indicated by a $\times$ in each panel. The
Br$\gamma$ image was obtained using the Palomar Integral field
Spectrograph (PIFS, Murphy et  al. 1999). 
 
 Figure~6b  A contour map of NGC 6090 at 12.5~$\mu$m is overlaid on
the corresponding  grayscale image at 2.2~$\mu$m (left) and 8.4 GHz
image (right).

 Figure~6c  The spectral energy distributions of the sources  in NGC
6090 are shown. The filled circles represent the integrated fluxes
taken from IRAS and 2MASS data.  The filled squares represent
photometry in a 6$''$ diameter beam centered on the eastern source. The
open squares and open triangles represent photometry in a 1.5$''$
diameter beam centered on the southern and northern components of the
eastern source. The open circles and filled squares marked with a W
present photometry in  1.5$''$ and 4$''$ diameter beams centered on the
western source.  The 2.2~$\mu$m photometric data are scaled from the
NICMOS image (S00).

\figcaption{}  Figure~7a A montage of contour plots is presented  of
Markarian 331 at 2.15~$\mu$m, 3.4~$\mu$m, 12.5~$\mu$m and 8.4~GHz. The
origin is  chosen to be at the bright nucleus in each panel and is
indicated  by a +  in each panel.  The 2.15~$\mu$m and
3.4~$\mu$m images are from the 200-inch Hale Telescope.

 Figure~7b  A contour map of Markarian 331 at 12.5~$\mu$m is overlaid
on the corresponding grayscale image of 8.4~GHz  emission.

 Figure~7c  The spectral energy distribution of sources in Markarian
331 are shown.  The filled circles represent the integrated fluxes
taken from IRAS and 2MASS data.  The aperture photometry in a 4$''$
diameter beam is represented by filled squares, the photometry in a
1.5$''$ diamater beam centered on the nucleus is represented by open
circles.  The photometry in the annulus between the 1.5$''$ diamater
and 4$''$ diameter is shown as open squares. 
\figcaption{} 

A montage of the normalized curves-of-growth (included flux density vs.
beam size) is shown  for the galaxies in this sample at 12.5~$\mu$m
(solid line), 2.2~$\mu$m (dashed line) and 1.1~$\mu$m (or 1.2~$\mu$m in
the case of Markarian 331, dotted line). The normalization to 1.0 is
taken from the IRAS 12 $\mu$m flux density for the 12.5~$\mu$m curve
and the integrated magnitudes of these galaxies at 2.2 and 1.2 $\mu$m
from the 2MASS all sky survey (Jarrett, private communication).  The
physical scale for each galaxy was taken from the data in Table 1.  No
correction has been applied for  the PSF size, so that in some cases,
e.g., NGC 3690~A and B and NGC 2623, the profile reflects the curve of
growth of the PSF.  In all cases the centers of the beams for the
curves-of-growth are at the peak surface brightness at 12.5~$\mu$m as
shown in figures 1a -- 7a. The same location is taken for the center of
the beams at all wavelengths.  In the case of NGC 3690 the center
location of NGC 3690 B is taken as the location of source B1.  The
effects of the 4$''$ chopping amplitude are seen in the 12.5~$\mu$m
curves for NGC~1614, NGC~2623, and IC~883.

\clearpage
\normalsize
\newpage

\normalsize


\clearpage

\begin{table}

\centerline {Table 1}

\caption{ Basic Properties of  Observed Galaxies}

\smallskip
\begin{tabular}{l c c c c c }
\tableline\tableline
Name & z & log L & log$M_{H_2}$ & Spectrum &  linear scale \\
     &  &   L$_{bol}$[L$_{\odot}$] &[M$_{\odot}$] & opt/ir  & pc/$''$ \\
\tableline

VV 114 = Arp 236 &    0.0200   & 11.62   & 10.44 & HII/PAH  &  400 \\

NGC 1614 = Arp 186 &    0.0159   & 11.62   & 10.03 & HII/...    &  320 \\

NGC 2623 = Arp 243 &    0.0185   & 11.54  & 9.77   & ..../PAH &  370 \\

NGC 3690/IC694 = Arp 299 & 0.0104   & 11.91   & 10.06 & HII/PAH+Sil &   210 \\

IC 883 = Arp 193 &    0.0233   & 11.51   & 9.87  &  LINER/PAH & 470 \\

NGC 6090 = Mrk 496 &    0.0293   & 11.51   & 10.15 & HII/...  & 590  \\

UGC 12812 = Mrk 331 &    0.0185   & 11.41   & 10.11 &  HII/PAH+Sil &  370 \\

\tableline

\end{tabular}

\end{table}

\clearpage

\vfill\eject
\begin{table}

\centerline {Table 2}

\caption{ Keck Mid-Infrared Flux Densities for Starburst Galaxies\tablenotemark{a} }

\smallskip
\begin{tabular}{l c c c c c c  c c }
\tableline\tableline
Object & 7.9$\mu$m & 8.8$\mu$m &  9.7$\mu$m &  10.3 $\mu$m & 11.7$\mu$m & 12.5$\mu$m &  12$\mu$m\tablenotemark{b} & 17.9$\mu$m \\
 & [mJy] &[mJy] &[mJy] &[mJy] &[mJy] &[mJy] &[mJy] &[mJy] \\
\tableline

VV114E & 457 & 245  &52$\pm$21& 70$\pm$10 &  237 &338 & 1100 & 452$\pm$61  \\

NGC 1614\tablenotemark{c} & \nodata & \nodata & \nodata & \nodata & 1030 &1220 & 1400  & \nodata \\

NGC 2623 & \nodata &\nodata &\nodata & \nodata &  151 & 201 & 240 &\nodata \\

NGC 3690 sum\tablenotemark{d} & \nodata &\nodata &\nodata & \nodata & 2390 & 4030 & 3900 & 12650 \\

IC 883& 335$\pm$34 & \nodata & \nodata & 34$\pm$11 & 130 & 301$\pm$46 & 260 & \nodata \\

NGC 6090 sum\tablenotemark{e} &\nodata &\nodata &\nodata & \nodata & 167 & 134 & 290 & 356$\pm$40 \\

Mrk 331 & \nodata &\nodata &\nodata & \nodata & 280 & 403 & 510 & 685 \\

\tableline
\tablenotetext{a}{The
Keck flux densities are in a 4$''$ diameter circular beam unless otherwise 
noted.
The uncertainties are omitted from 
the table unless
the statistical uncertainty in the photometry, based on the noise in
the sky, exceeds 10\%. The uncertainties quoted are statistical only;
photometric uncertainties are on the order of $\pm$10\%.} 
 \tablenotetext{b} {The 12 $\mu$m flux densities are from the IRAS data for these objects.  The effective beam size is 2$'$ for the IRAS measurements. The
uncertainties in the IRAS data are 5-10\%.}
 \tablenotetext {c} {At 24.5~$\mu$m a flux density of 7.0 Jy was 
measured} 
 \tablenotetext{d} {The flux densities are the sum of the flux densities 
measured in four separate 4$''$ diameter beams centered on 
NGC~3690 A, NGC~3690 B1, NGC~3690 C, and NGC~3690 C$'$, and a 2$''$ 
diameter circular beam centered at NGC~3690 B2.}
 \tablenotetext{e} {The flux densities in a 6$''$ diameter beam 
centered on NGC~6090E$_{S}$ and a 4$''$ beam centered on NGC~6090W}
\end{tabular}

\end{table}

\clearpage

\vfill\eject

\begin{table}
\small
\scriptsize

\centerline {Table 3}

\caption{ Sizes and surface brightnesses in Starburst Galaxies }

\smallskip
\begin{tabular}{l c c c c c c  c r }
\tableline\tableline
Object &\multicolumn{2}{c} {size of mid-ir source} &
$\frac {f_{12.5\mu m}(Keck)}{f_{12\mu m}(IRAS)}$ &
T$_b$(12$\mu$m) &T$_b$(60$\mu$m) &
T$_c$(60$\mu$m) & L$_{\it ir}$[L$_{\odot}$] &
{$\frac{L}{A}$[L$_{\odot}$Kpc$^{-2}$]}  \\
        & $''$ & pc &    &K &K &K &  & \\

\tableline

VV114E sw& 0.37& 150 & 0.30 &94 &57 &54 & $1.2\times 10^{11}$ & \\

  & $>0.41$& $>165$ & 0.30 &92 &54 &54 & $1.2\times 10^{11}$ & $<5.6\times
10^{12}$\\

VV114E ne& 0.29& 120 & 0.12 &91 &51 &54 & $5.0\times 10^{10}$ & $4\times
10^{12}$\\

NGC 1614 & 1.70& 550 & 0.87 &84 &41 &62 & $3.6\times 10^{11}$ & $1.5\times
10^{12}$\\

NGC 2623 & $ 0.7 \times < 0.3$ & $ 260 \times < 110 $ & 0.85 &$>$89 &$>$65
&64 & $3.0 \times 10^{11}$ &  \\

  & $>0.52$ & $> 192 $ & 0.85 &88 &64 &64 & $3.0 \times 10^{11}$ & $<1.0
\times 10^{13}$\\

NGC 3690 A&$<0.60$ &$<125$ & 0.32 &$>$100 &69 &66 & $2.6\times 10^{11}$ & \\
  &$>0.65$ &$>137$ & 0.32 &98 &66 &66 & $2.6\times 10^{11}$ &$<1.8\times
10^{13}$ \\
NGC 3690 C$'$&$<0.60$  &$<125$ & 0.04 &$>$85 &$>$44 &66 & $3.2\times
10^{10}$ &$>2.6\times 10^{12}$ \\
NGC 3690 C&1.1 &230 & 0.12 &79 &42 &66 & $1.0\times 10^{11}$ &$2.4\times
10^{12}$ \\
NGC 3690 B1&$<0.60$ &$<125$ & 0.51 &$>$104 &$>$80 &66 & $4.0\times
10^{11}$ & \\

  &$>0.84$ &$>176$ & 0.51 &98 &66 &66 & $4.0\times 10^{11}$ &$<1.6\times
10^{13}$ \\
NGC 3690 B2&$<0.6$ & $<125$ & 0.015 &$>$78 &$>$37 &66 &  $1.2\times
10^{10}$ &$>1.0\times 10^{12}$ \\

IC 883& $1.5 \times 0.4 $ & $700 \times 190$ & 1.00 &84 &20 &46 & 
$3.2\times 10^{11}$ & $2.0\times 10^{12}$\\

NGC 6090E S& $1.5 \times 1.0$&$ 890 \times 590$ & 0.30 &73 &30 &51 & 
$9.7\times 10^{10}$ & $2.4\times 10^{11}$\\

NGC 6090E N& 1.0 &590 & 0.12 &71 &28 &51 & $3.9\times 10^{10}$ & $1.4\times 
10^{11}$\\

NGC 6090W & $<$0.6 &$<$350 & 0.04 &71 &$>$29 &51 & $1.3\times 10^{10}$ & $> 
1.3\times 10^{11}$\\

Mrk 331 n&($<$0.6)  & $<$225 & 0.2  &$>$81 &$>$41 &58 & $5\times 10^{10}$ & 
$> 1.3 \times 10^{12}$ \\
Mrk 331 r&$4 \times 2.7$  & $1500 \times 1000$ &  0.63 &71 &30 &58 & 
$1.6\times 10^{11}$ & $1.4\times 10^{11}$\\

\tableline

\end{tabular}

\end{table}

\vfill\eject
\clearpage

\begin{table}
\small

\caption{\bf Surface Brightnesses of Infrared Starburst Galaxies}

\bigskip
\begin{tabular}{l c c c}   
\tableline\tableline
Object & Type & Infrared Luminosity & Surface Brightness  \\

 &      &   L$_{bol}$[L$_{\odot}$] & [L$_{\odot}$Kpc$^{-2}$]\\

\tableline

ORION  &  HII Region & 1$\times 10^6$ & $2 \times 10^{12}$ \\

M 82  &  Local Starburst &  3$\times 10^{10}$ & 2$\times 10^{11}$\\
\tableline
NGC 6090 & Starburst & 3 $\times 10^{11}$ & 2 $\times 10^{11}$ \\

NGC 1614 & Starburst &  4 $\times 10^{11}$ & 1.5 $\times 10^{12}$ \\

Mrk 331 & Starburst & 2.5 $\times 10^{11}$ & $\sim 2 \times 10^{12}$ \\

IC 883 & Starburst & 3 $\times 10^{11}$ & 2 $\times 10^{12}$ \\

VV 114 & Starburst & 4 $\times 10^{11}$ & $\sim 5 \times 10^{12}$ \\

NGC 2623 & Starburst &  3 $\times 10^{11}$ & $\sim 10^{13}$ \\

NGC 3690 & Starburst & 8  $\times 10^{11}$ & $\sim 10^{13}$ \\

\tableline

IRAS17208& ULIRG & 3 $\times 10^{12}$  & $1.2\times 10^{12}$ \\

Mrk 273 &  ULIRG & 1.3$\times 10^{12}$ & $>2.2\times 10^{13}$ \\
 
IRAS08572& ULIRG & 1.3$\times 10^{12}$ & $>2.8\times 10^{13}$  \\

Arp 220&  ULIRG & 1.5$\times 10^{12}$ & $6.0\times 10^{13}$ \\

\tableline

\end{tabular}

\end{table}

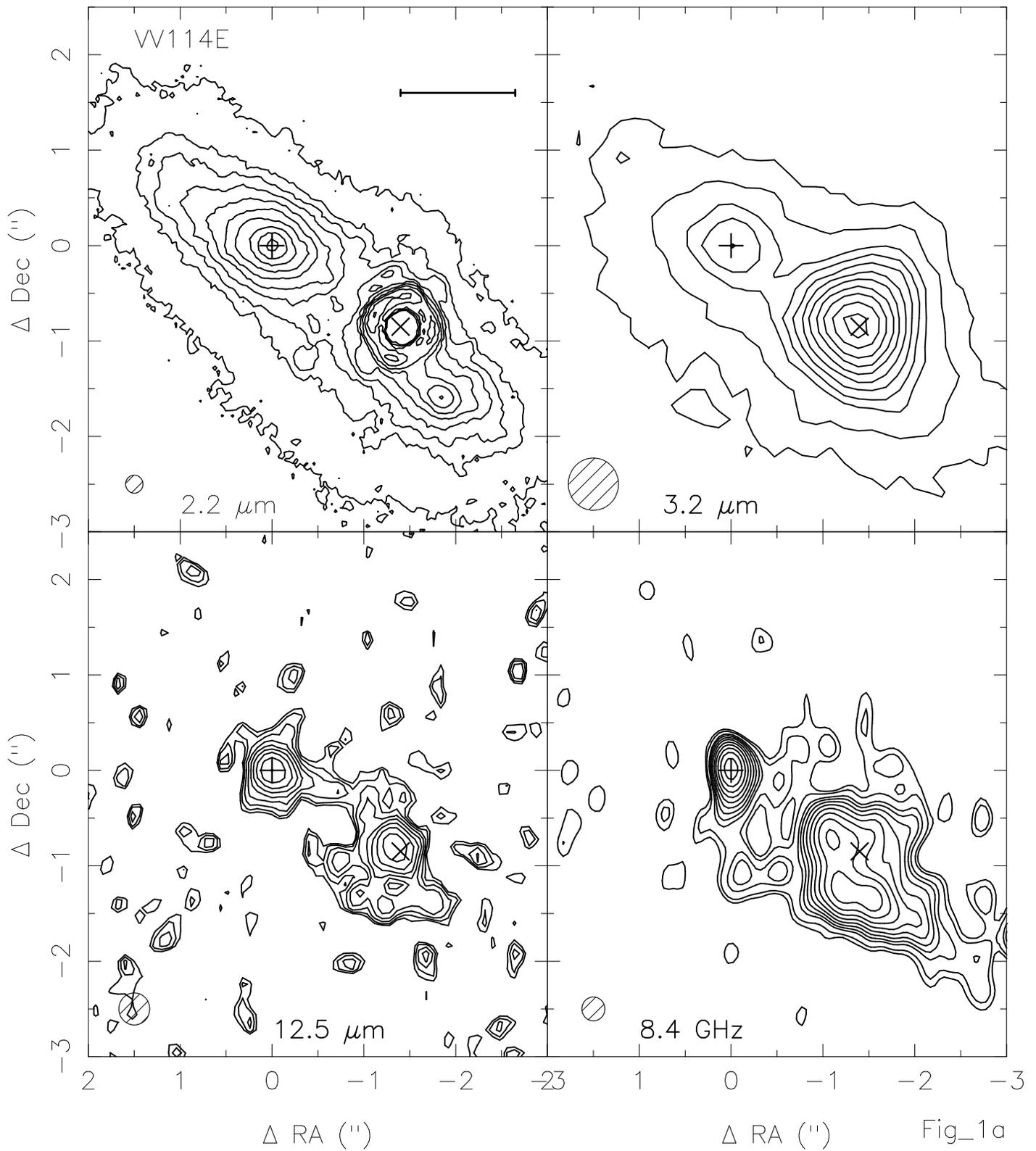
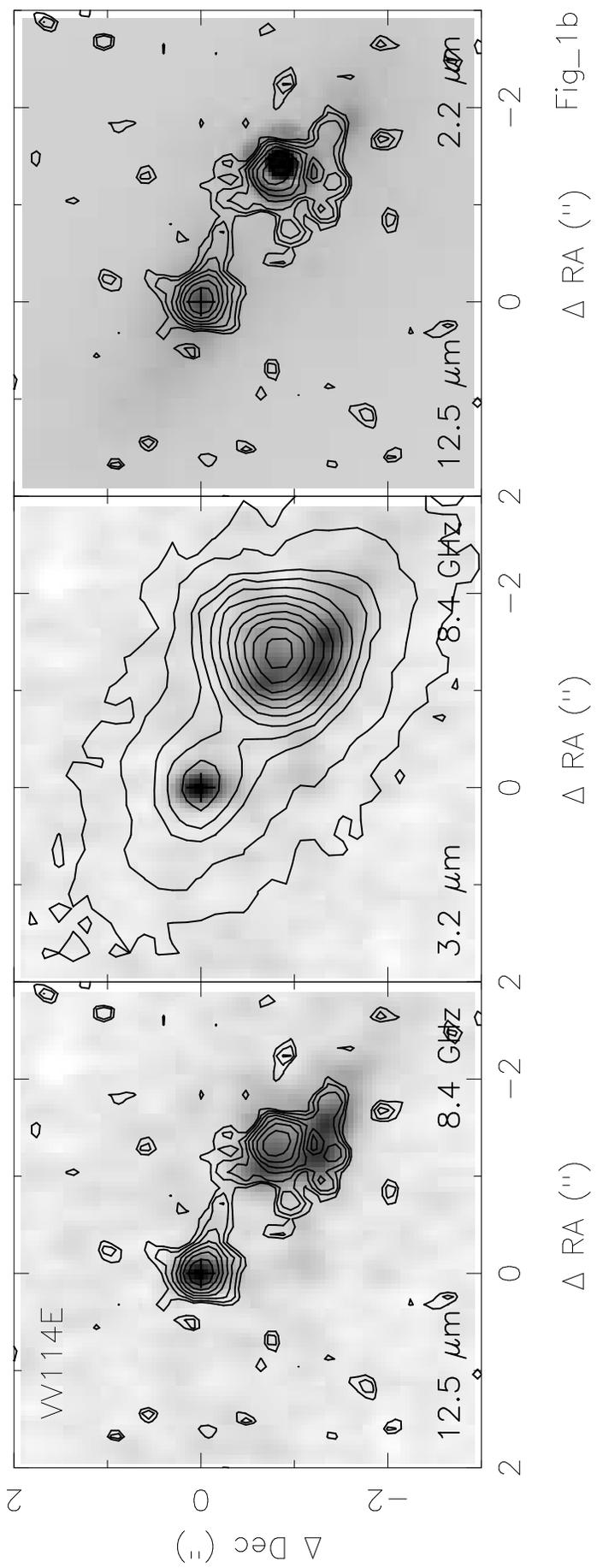
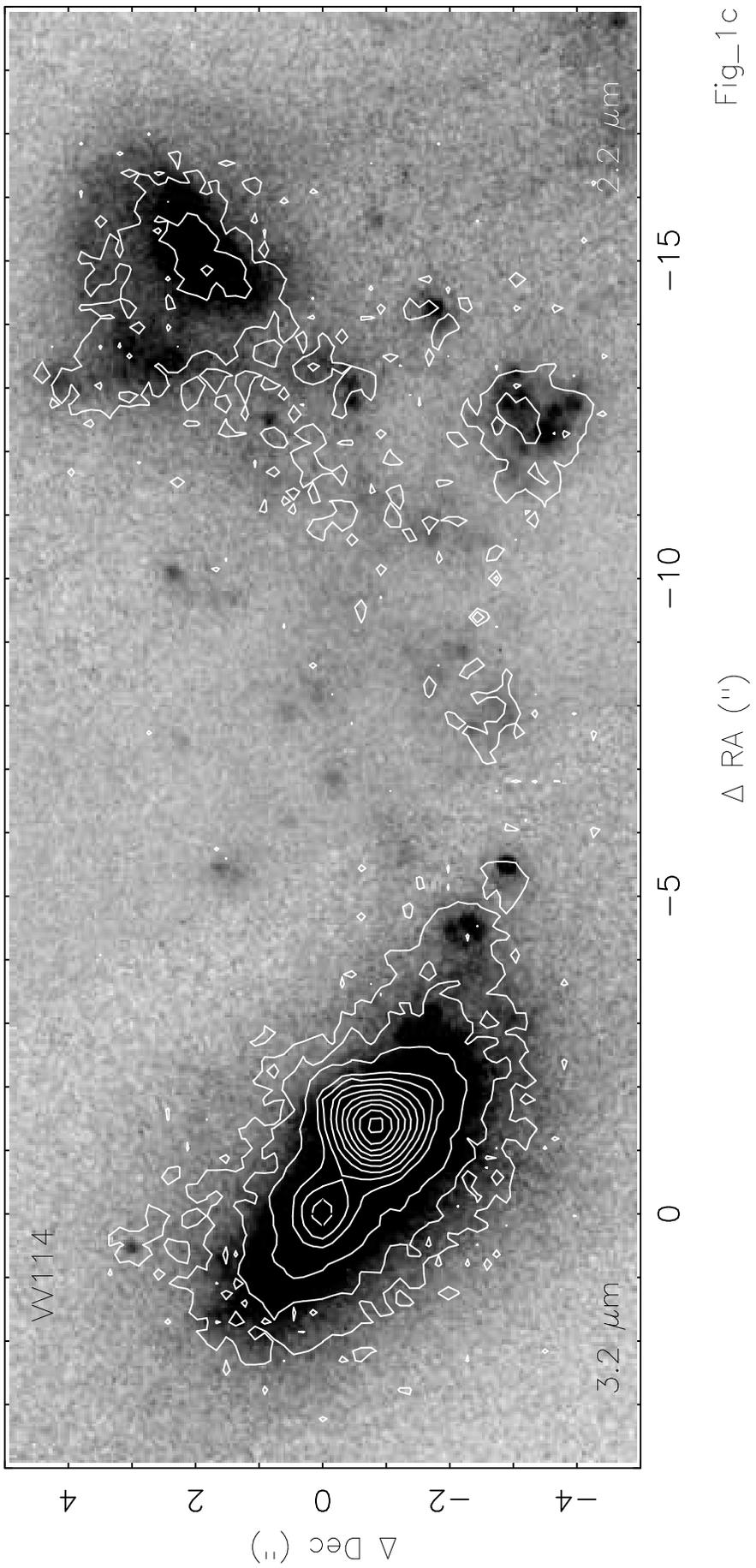
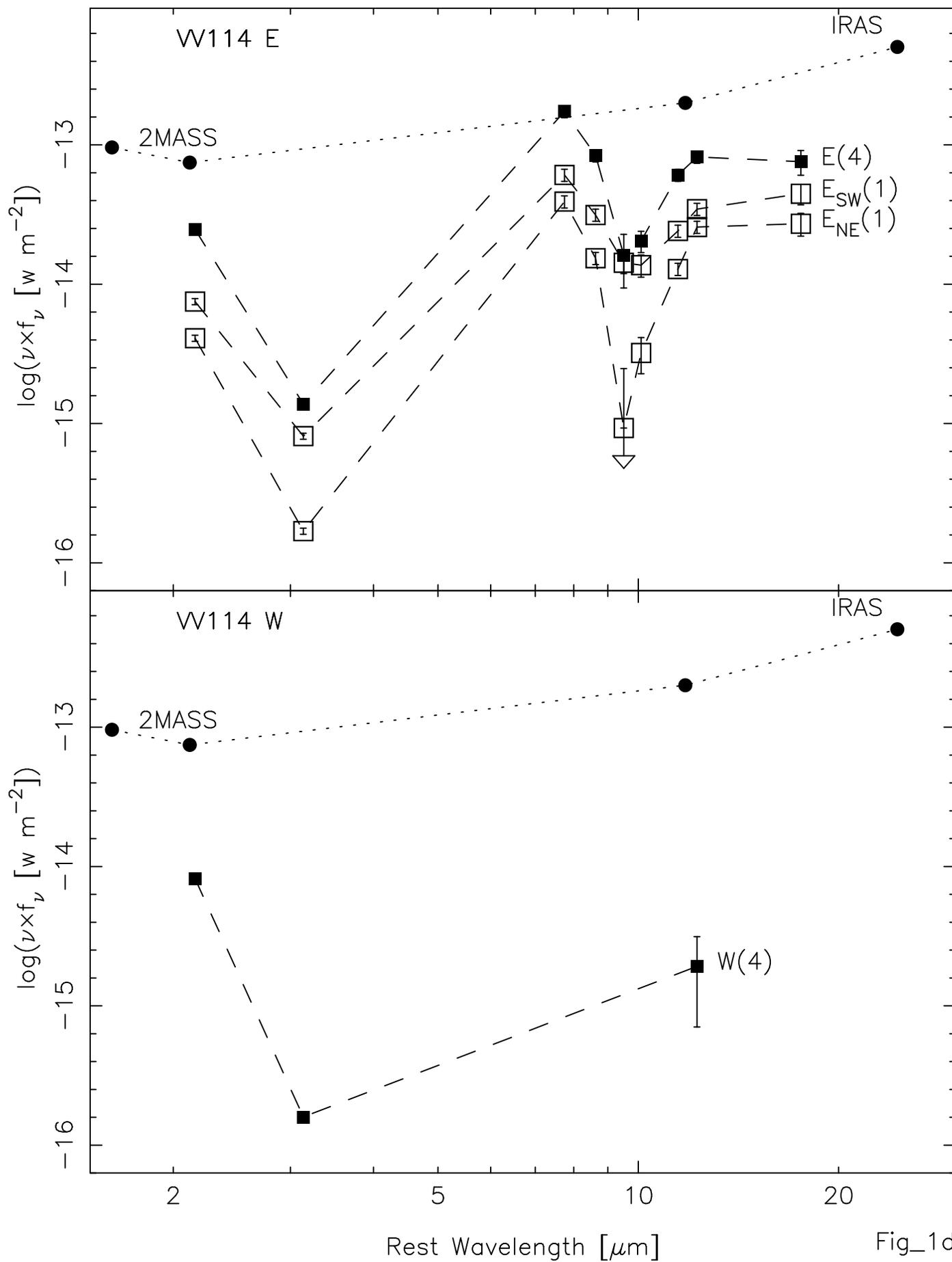
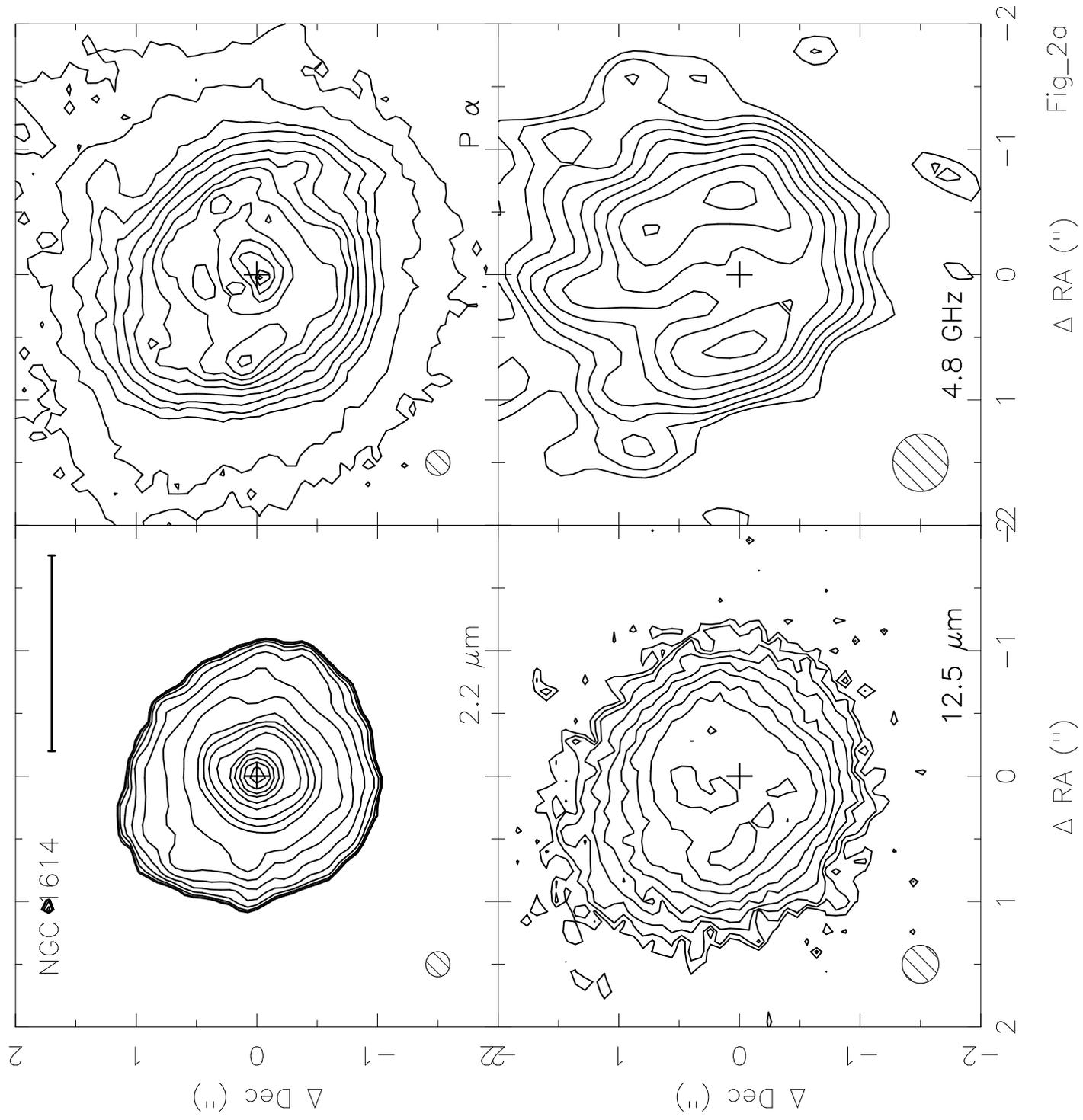
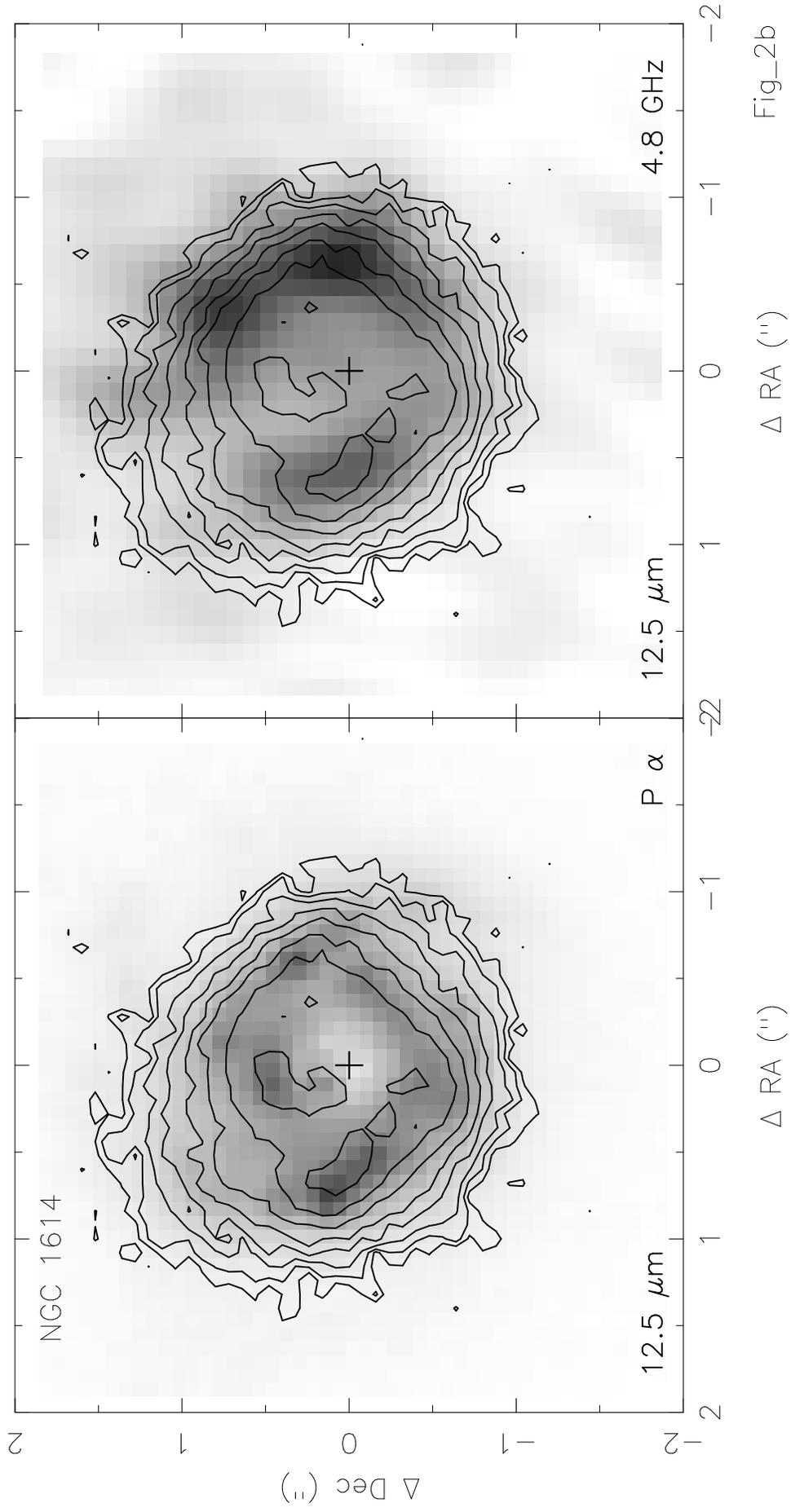
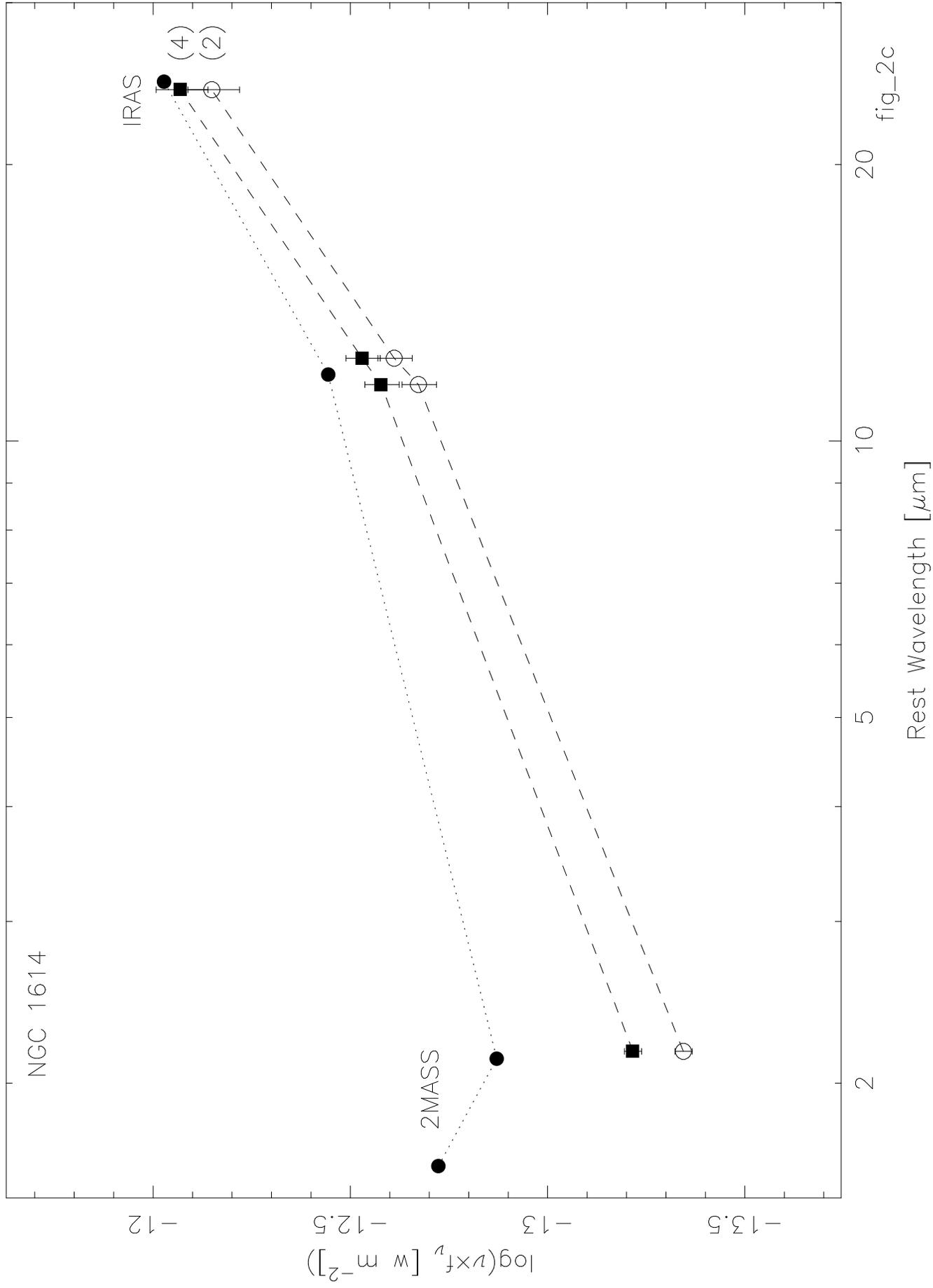
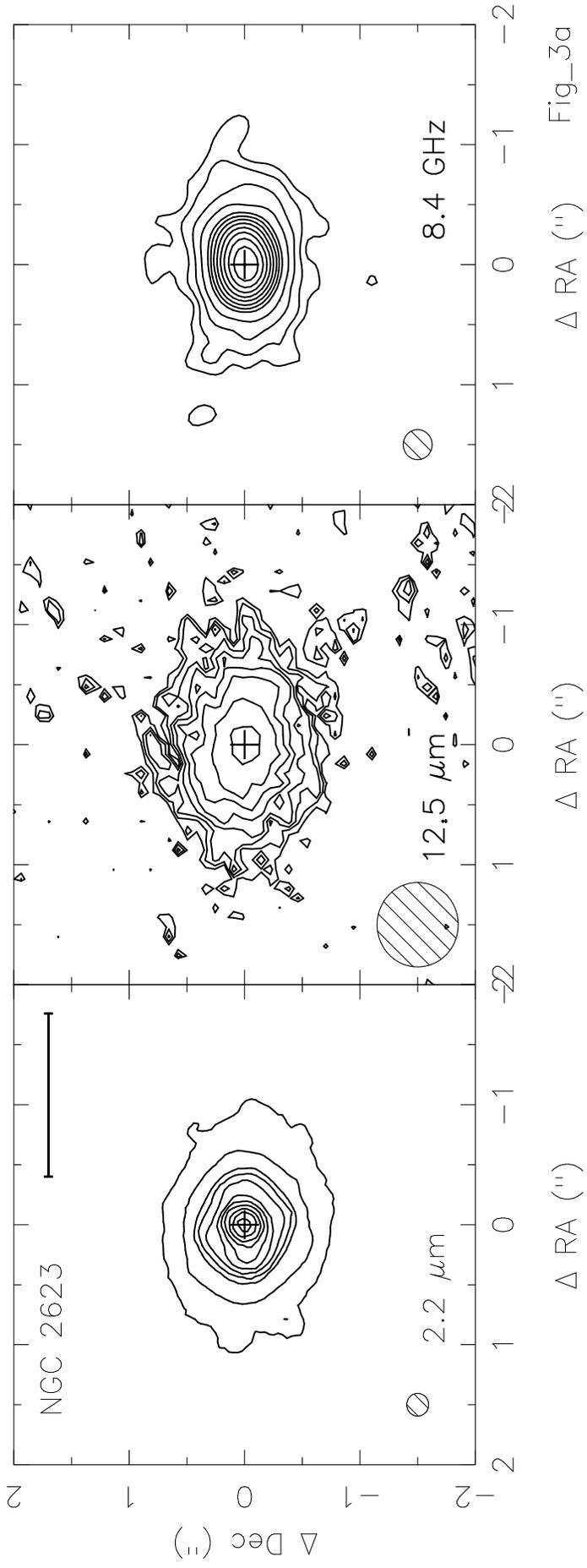
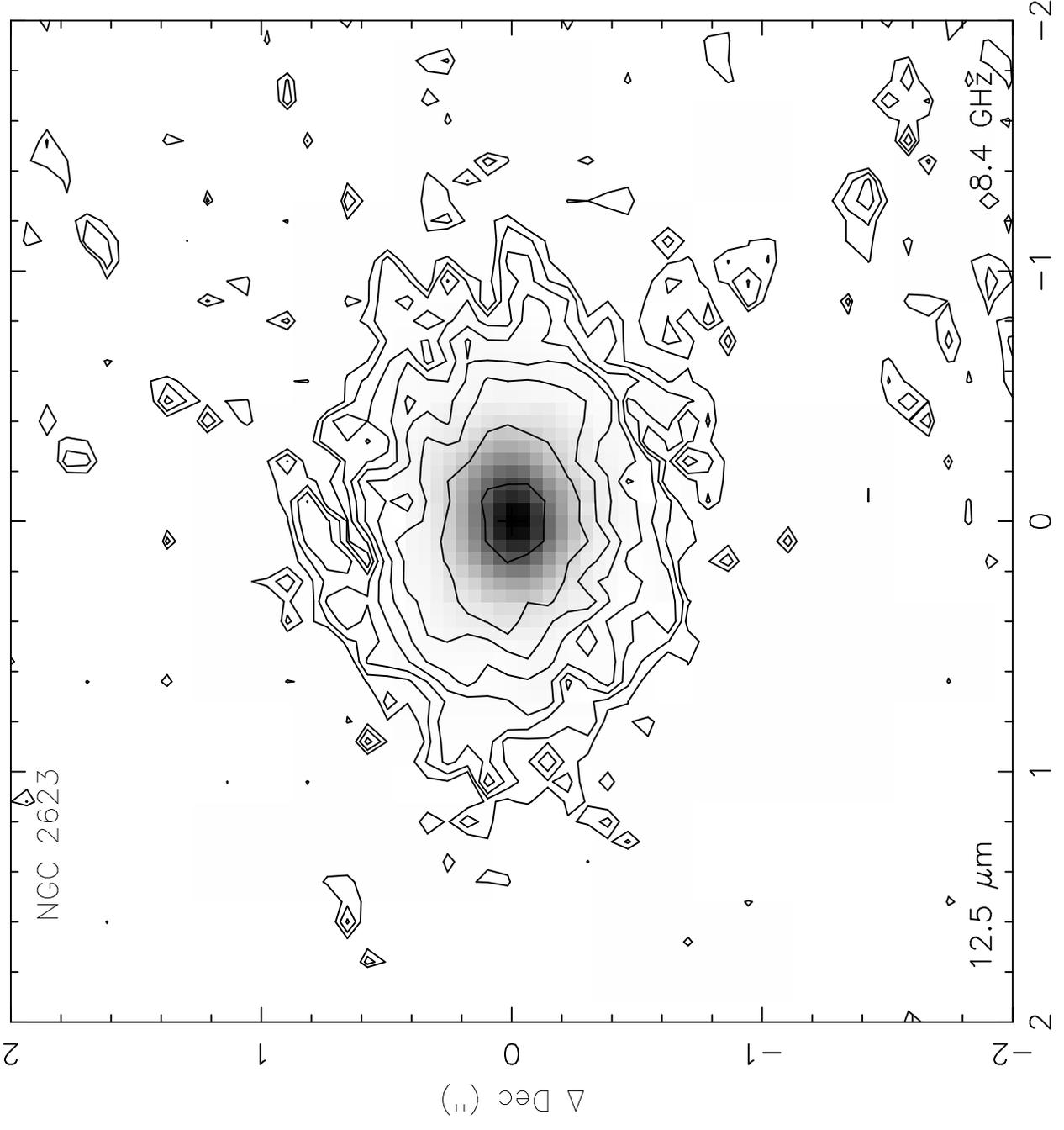
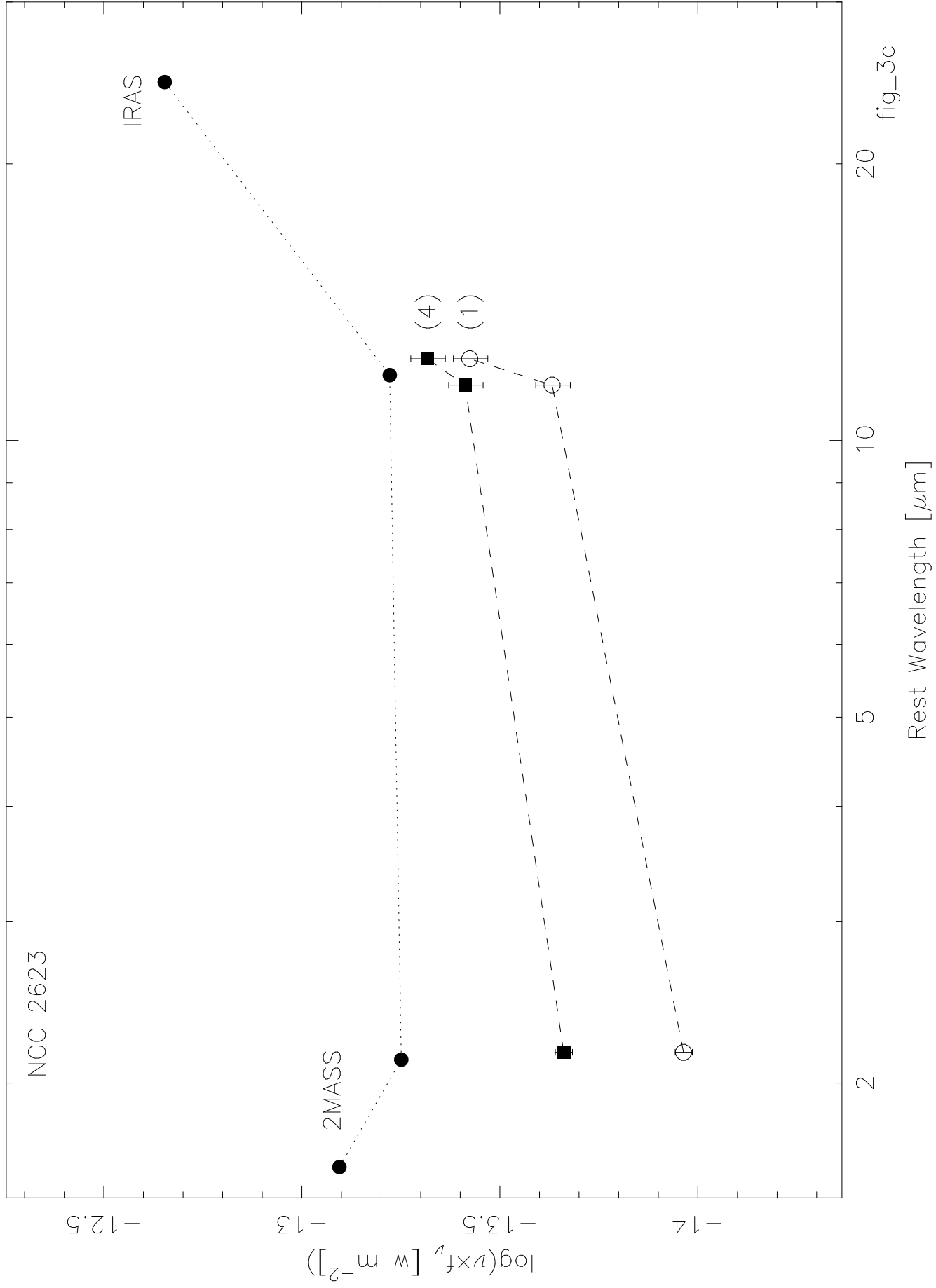
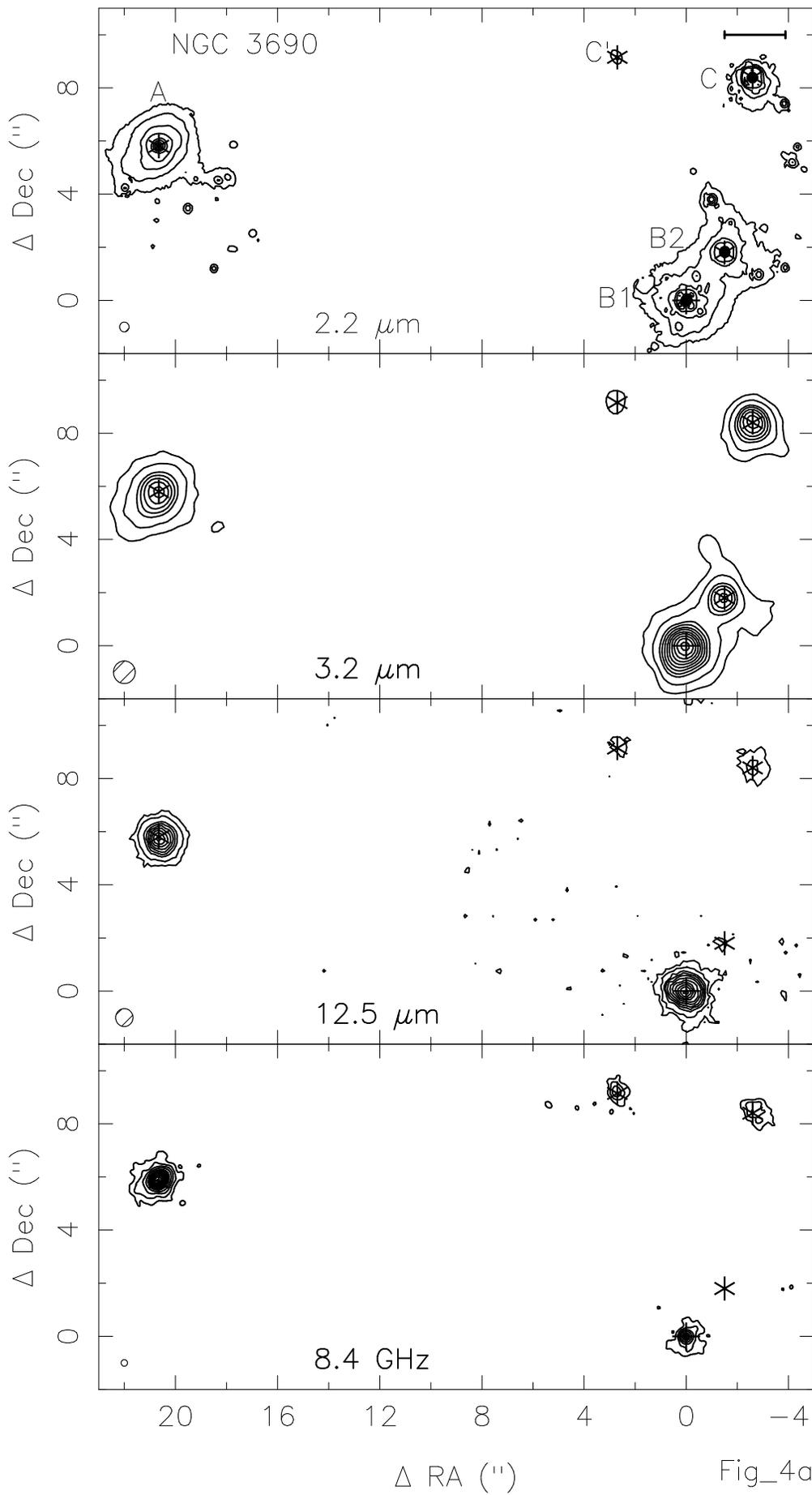
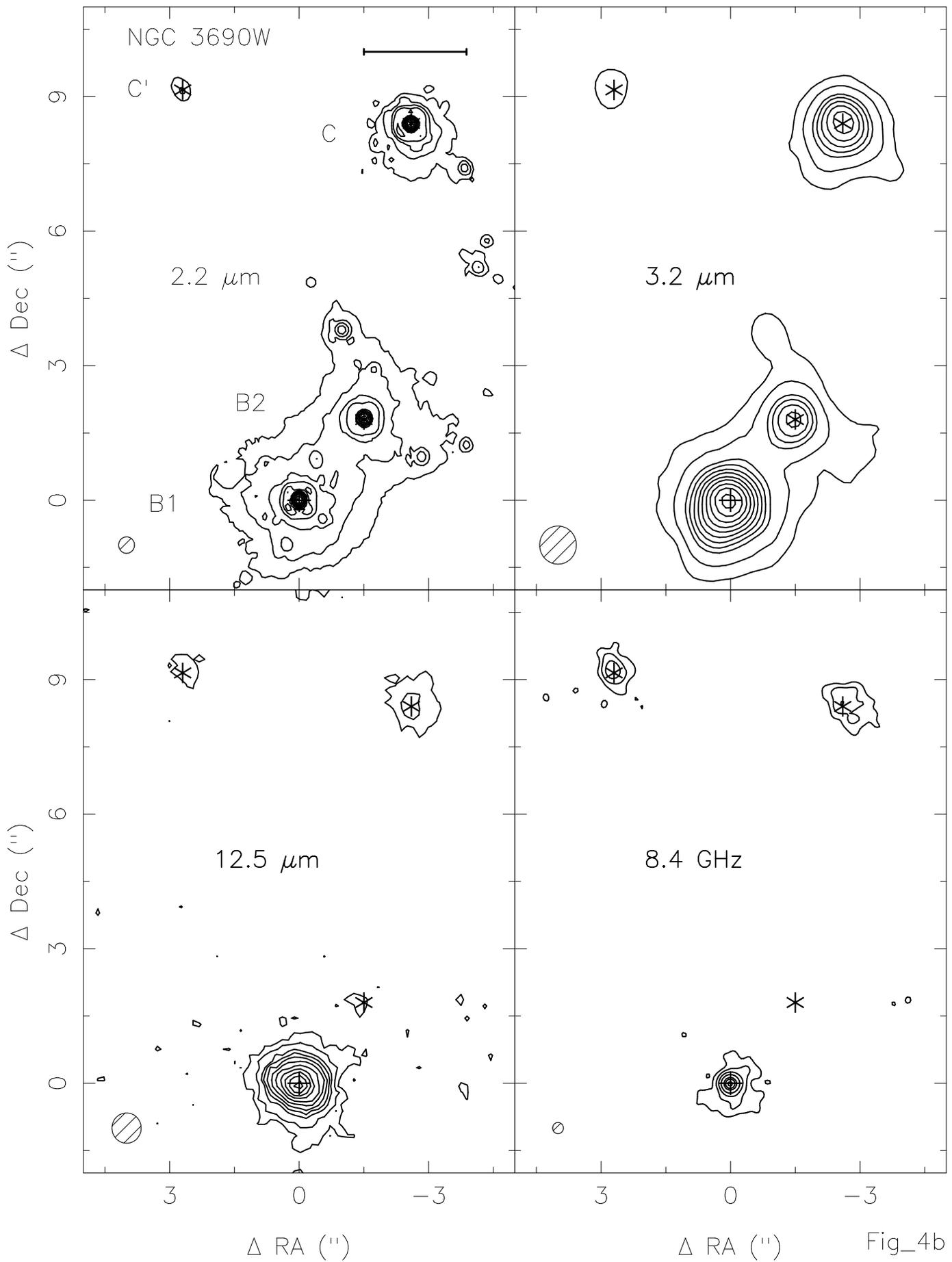
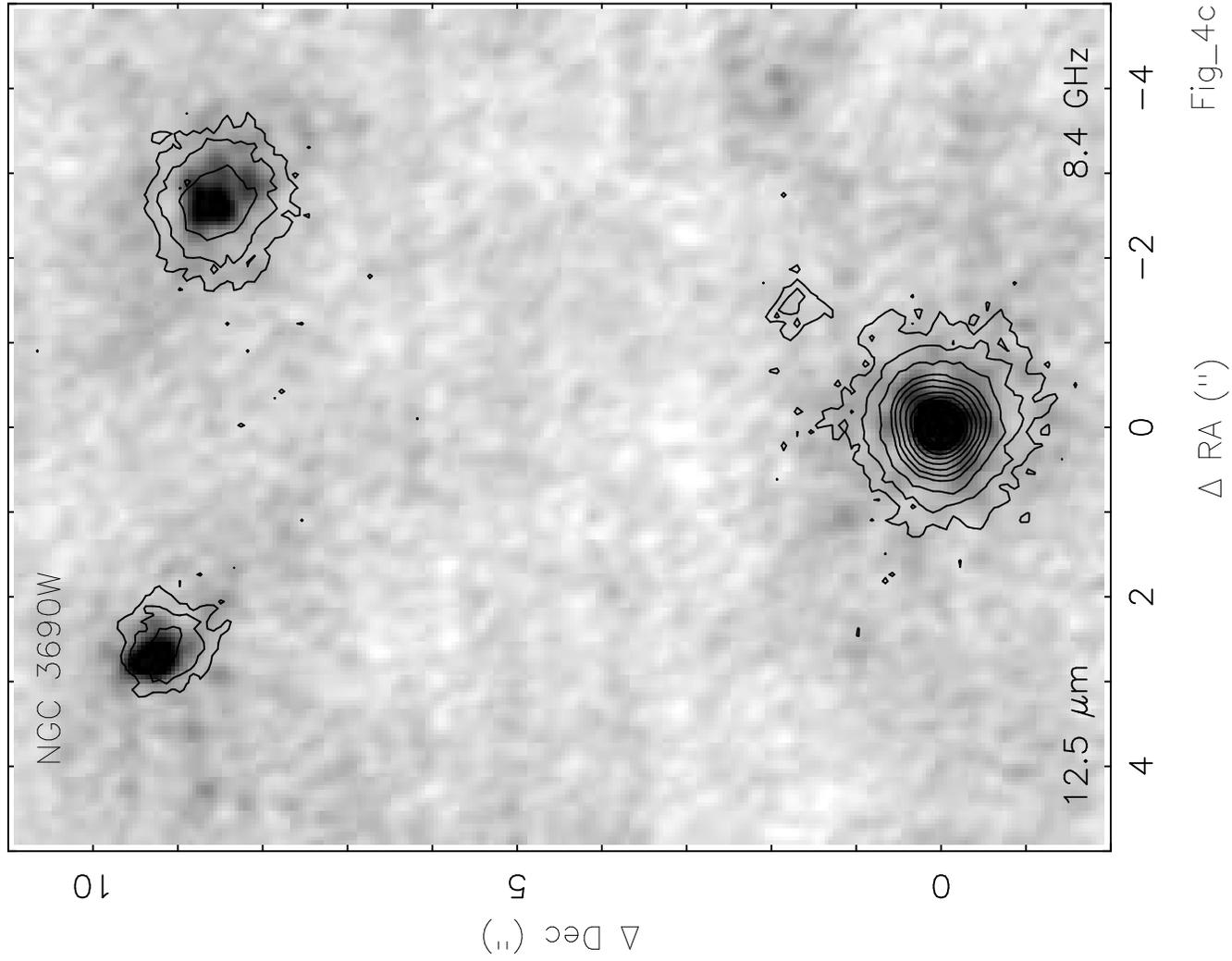
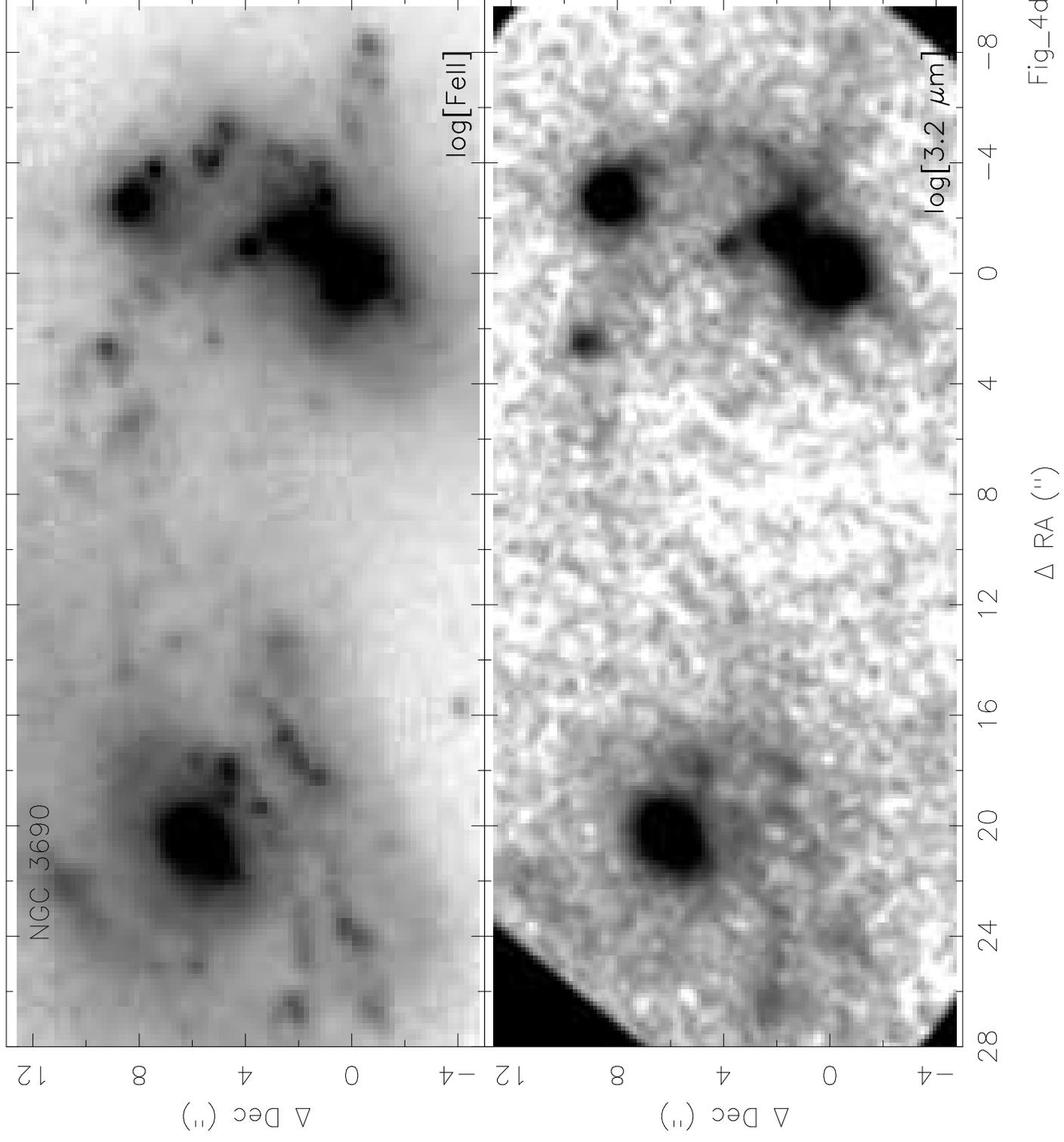
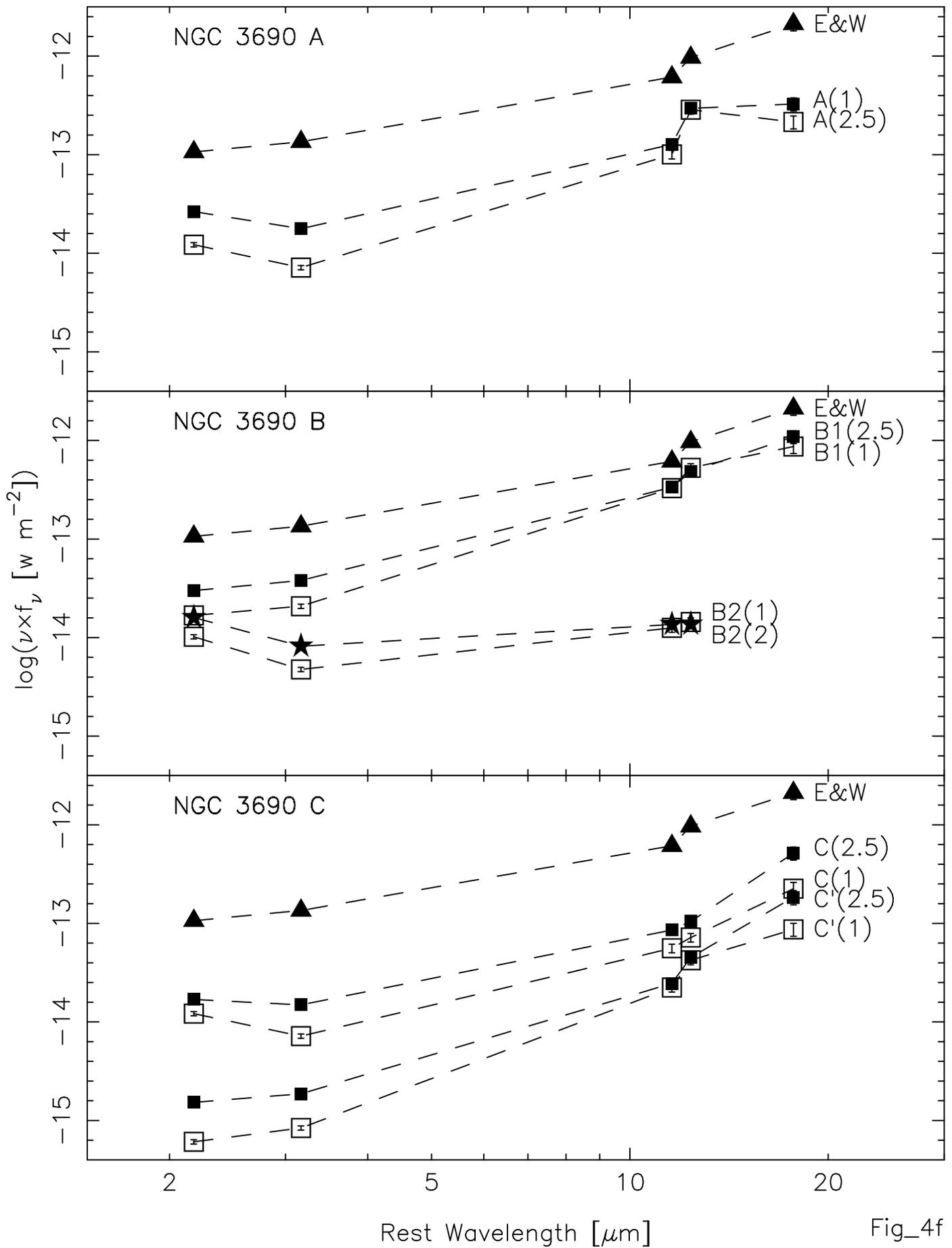
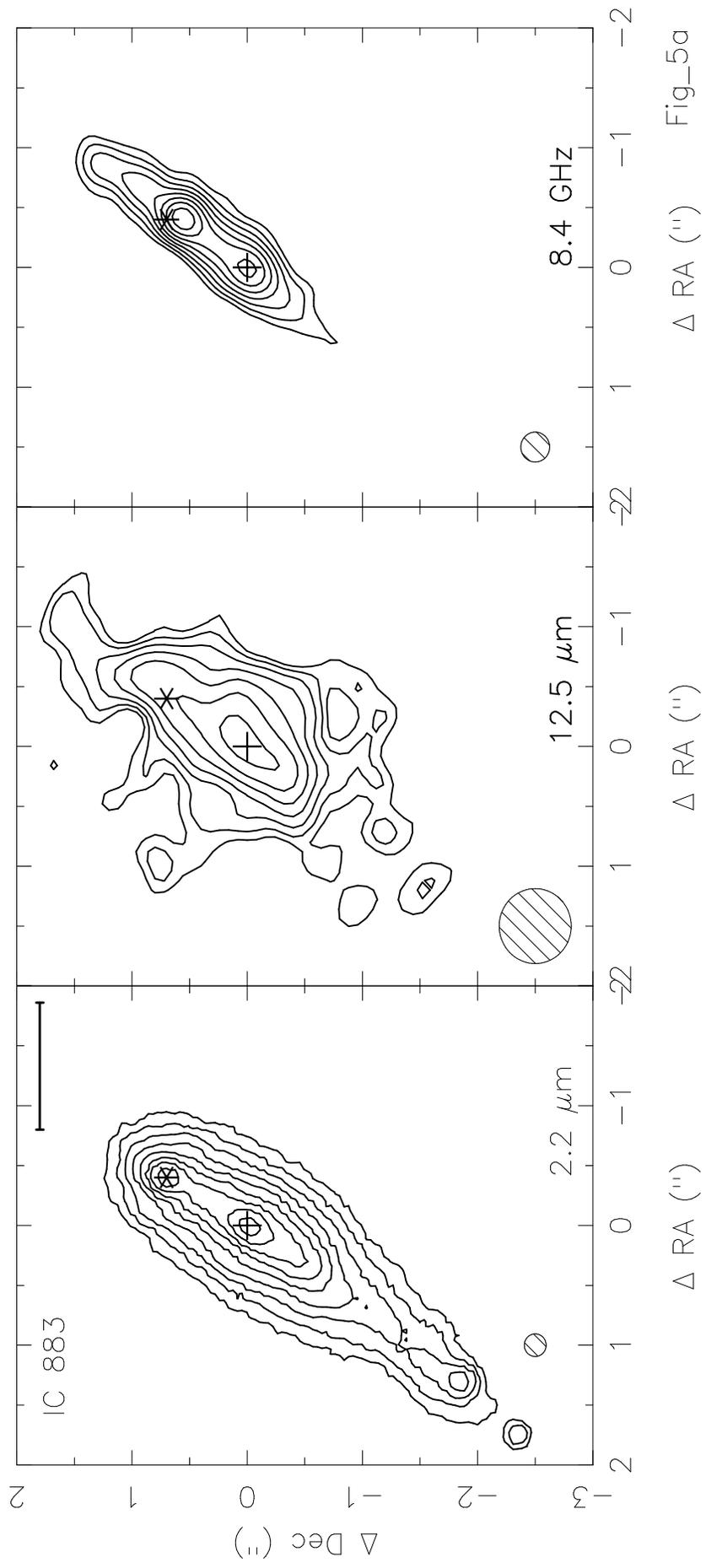
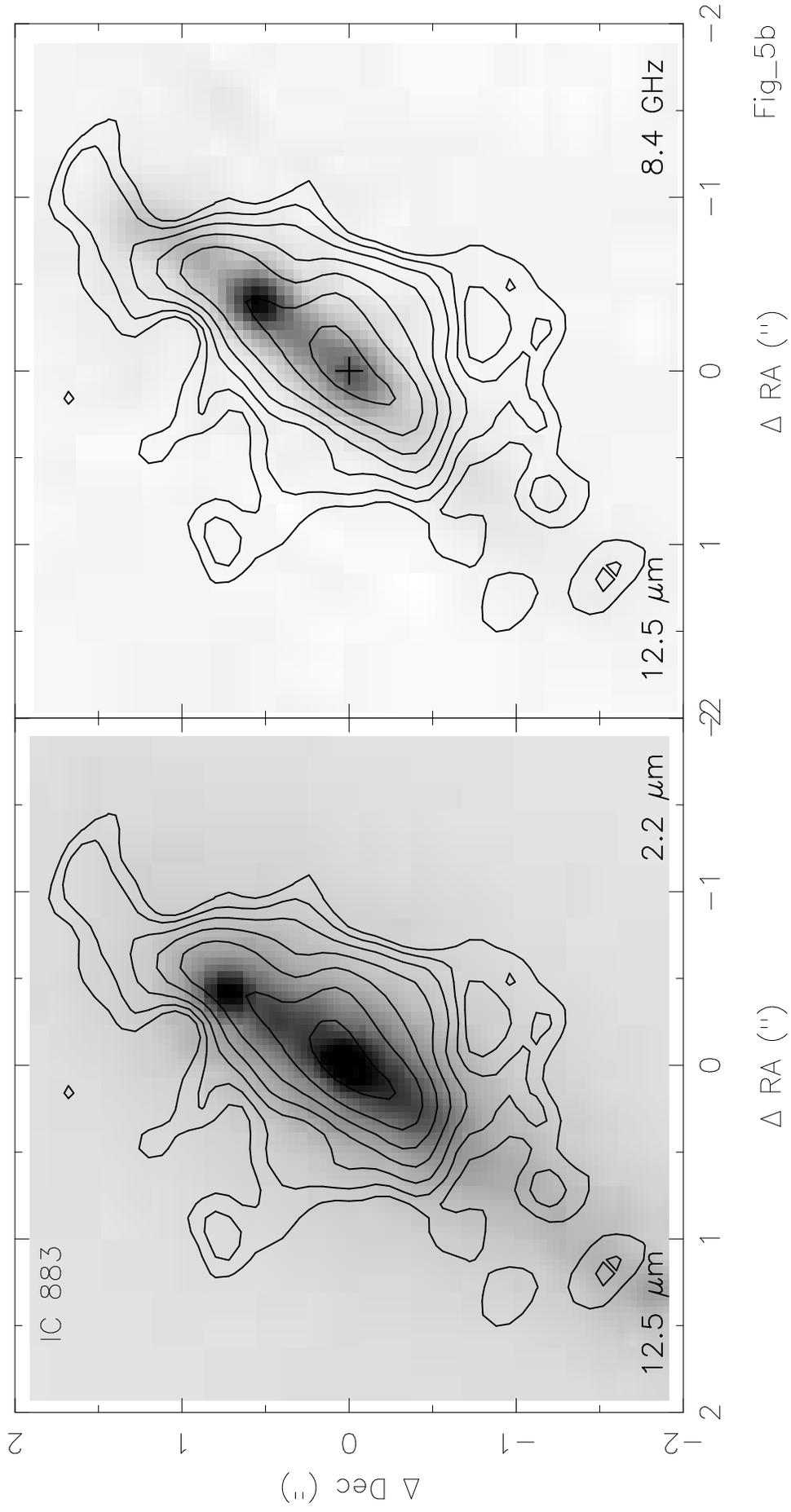
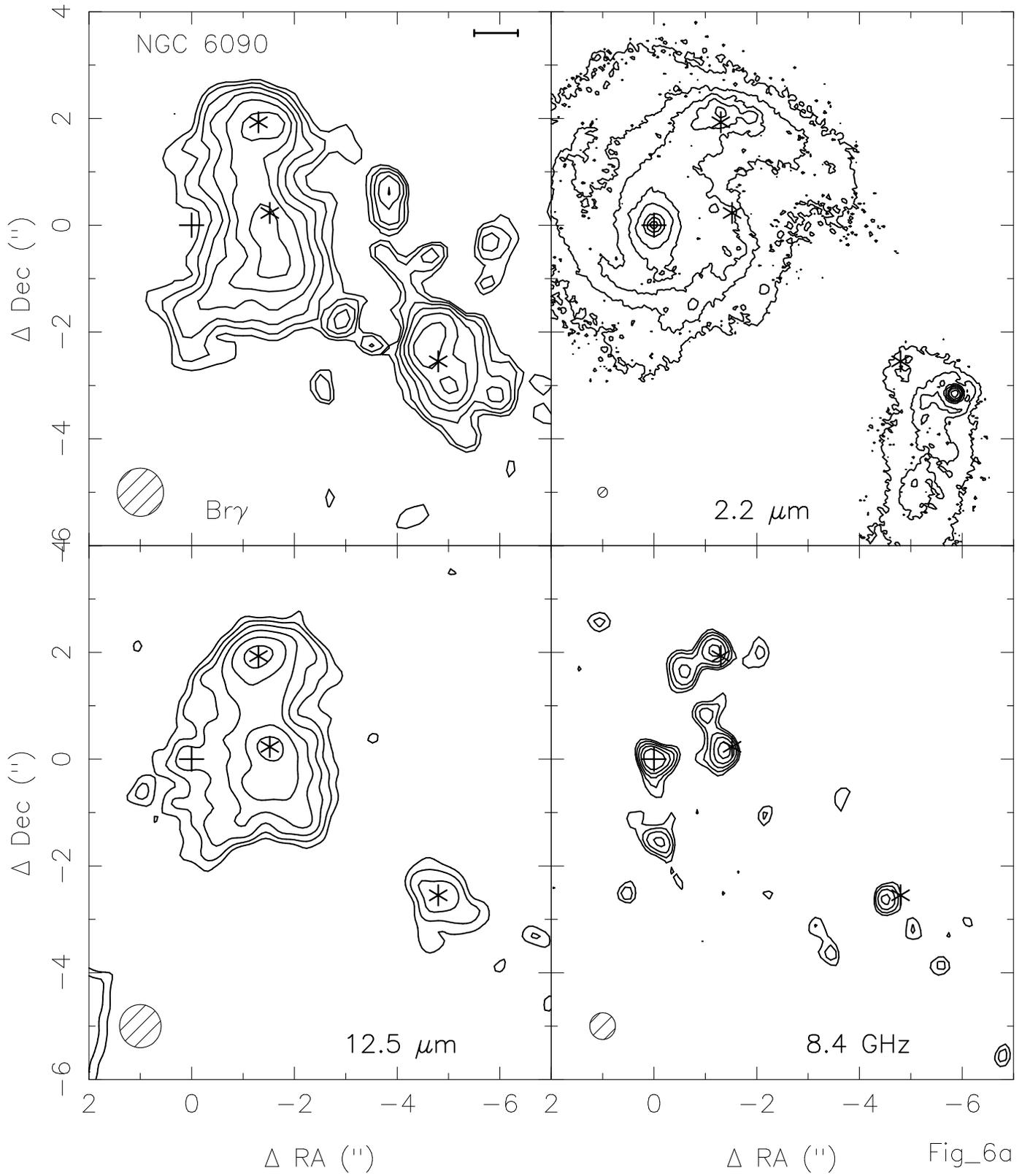
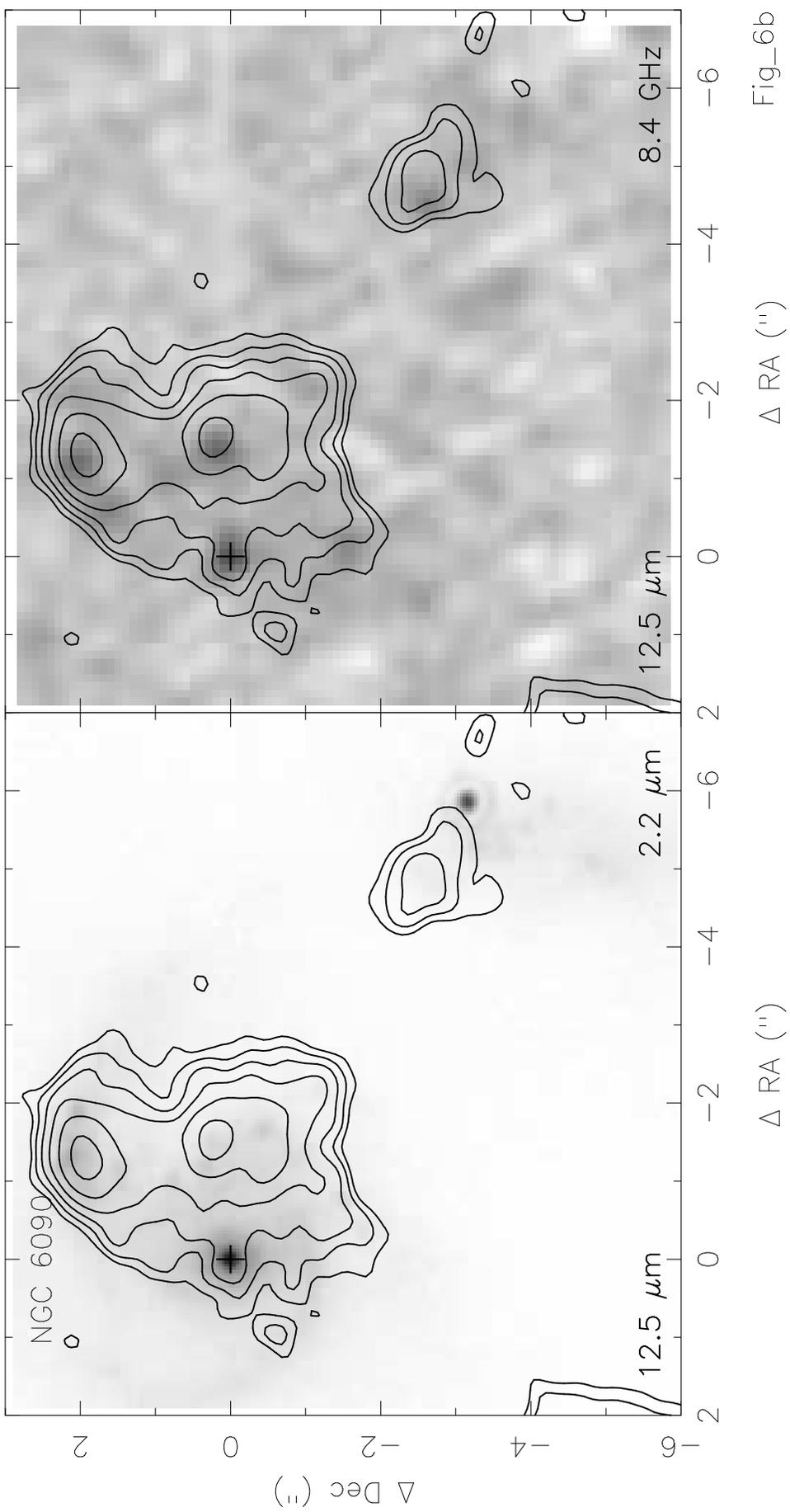
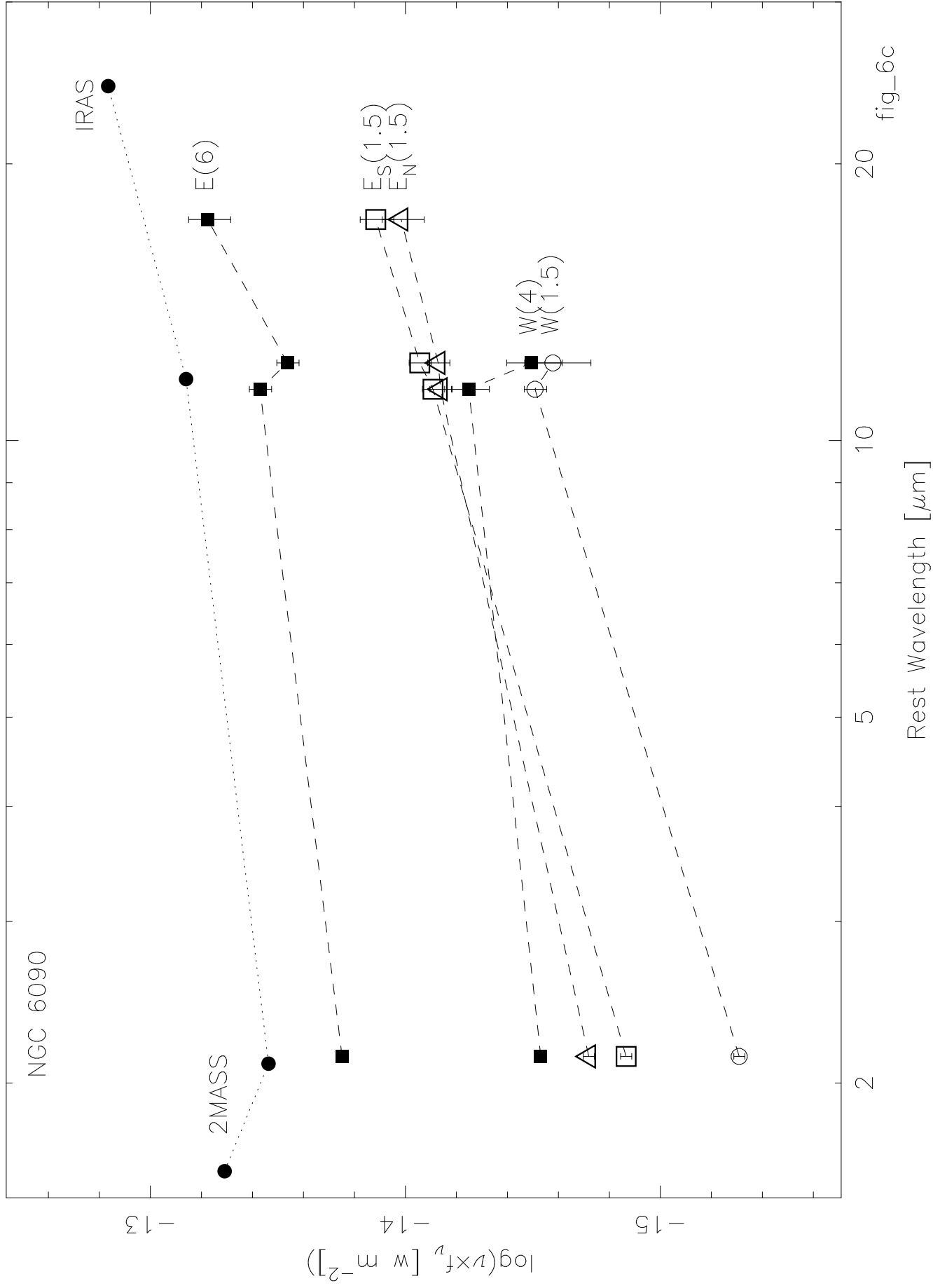
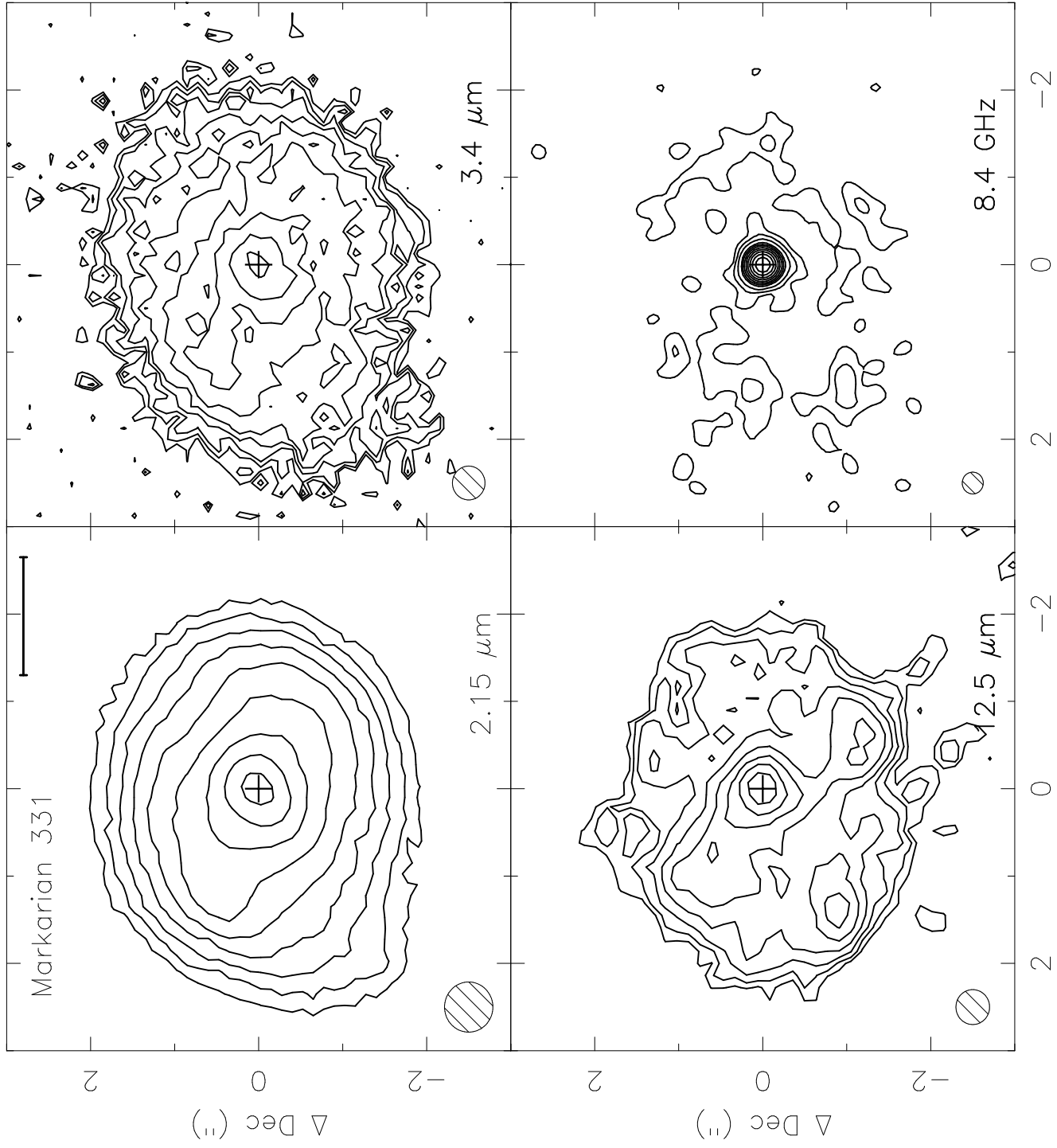
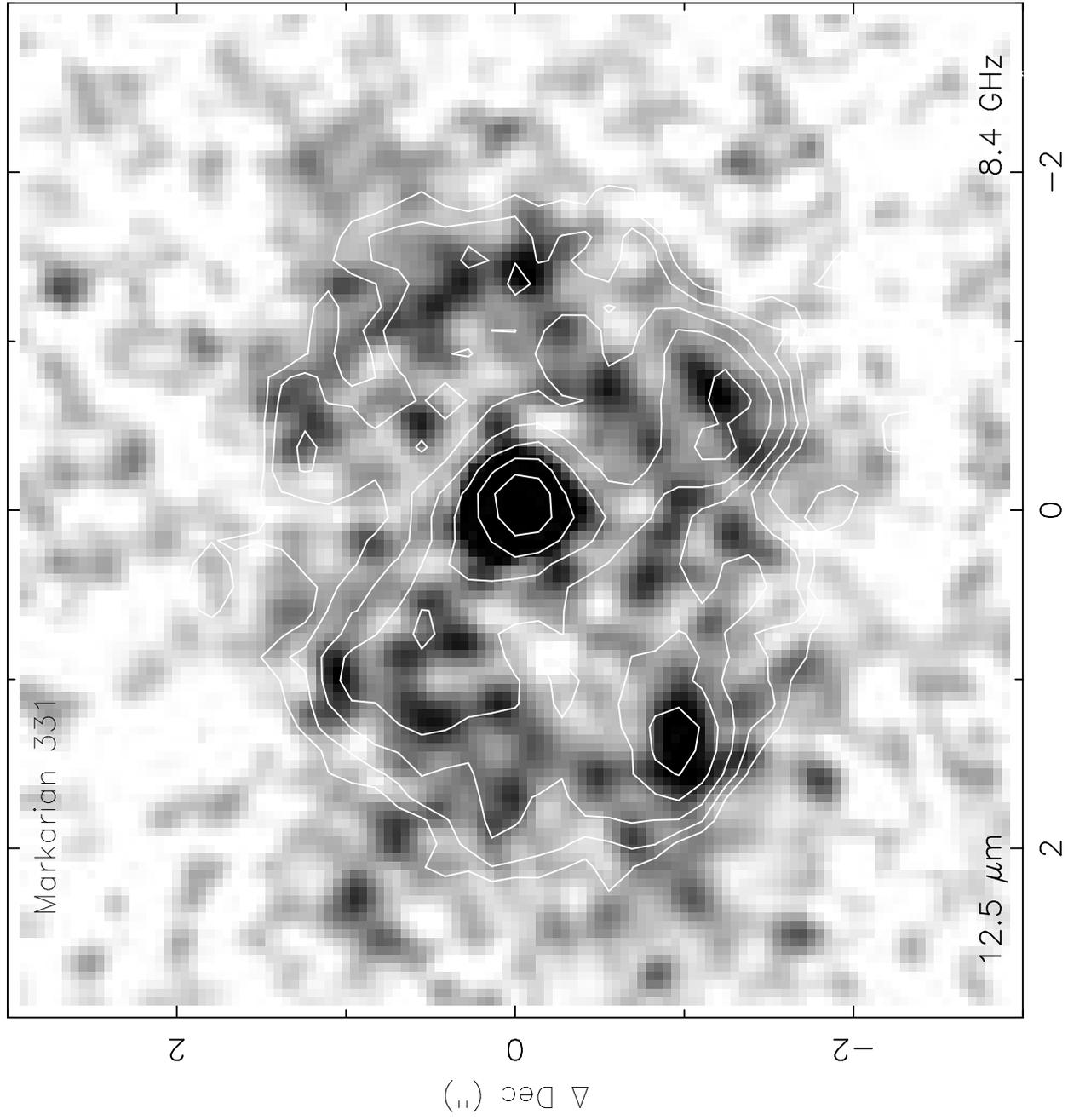
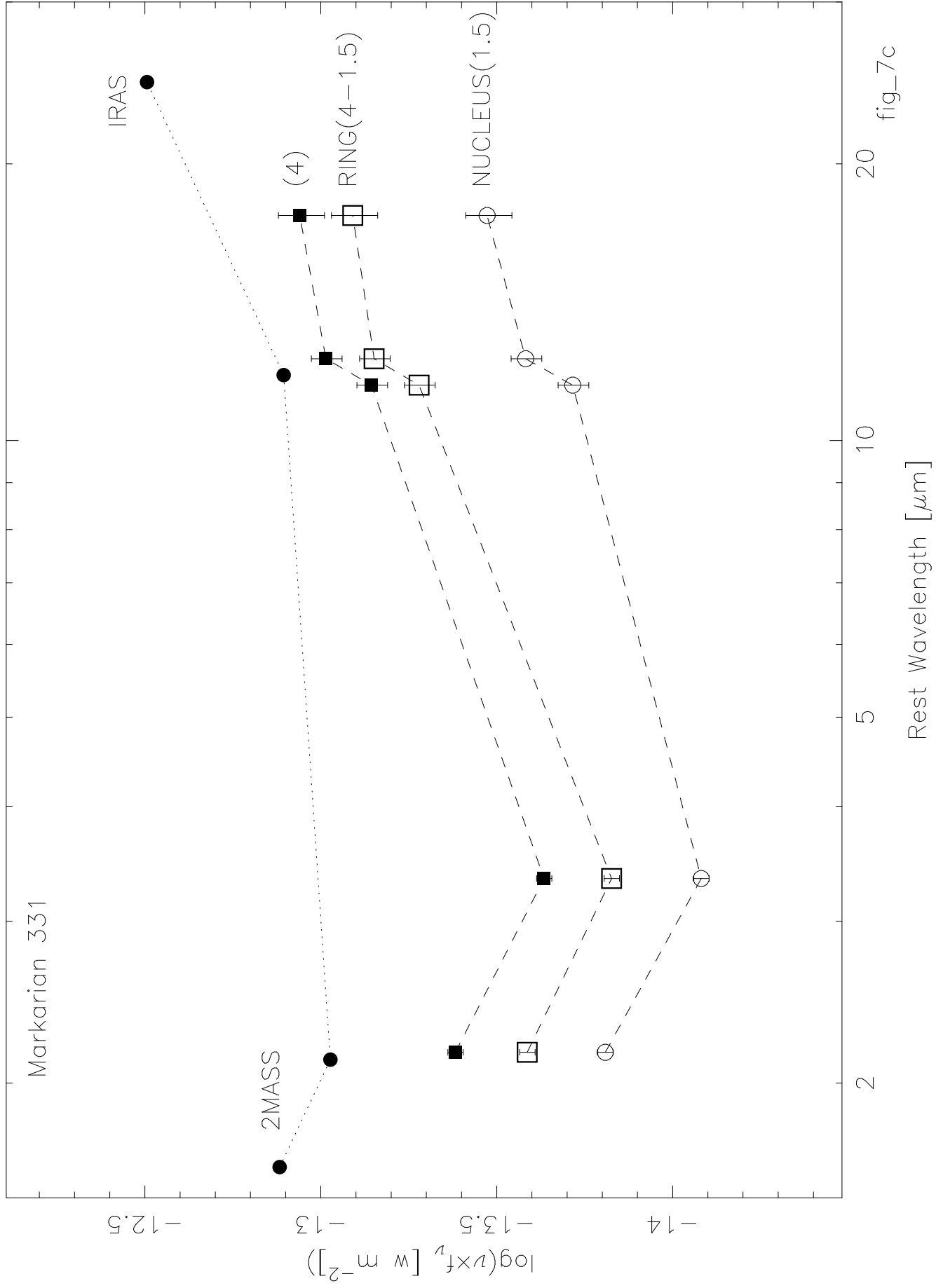
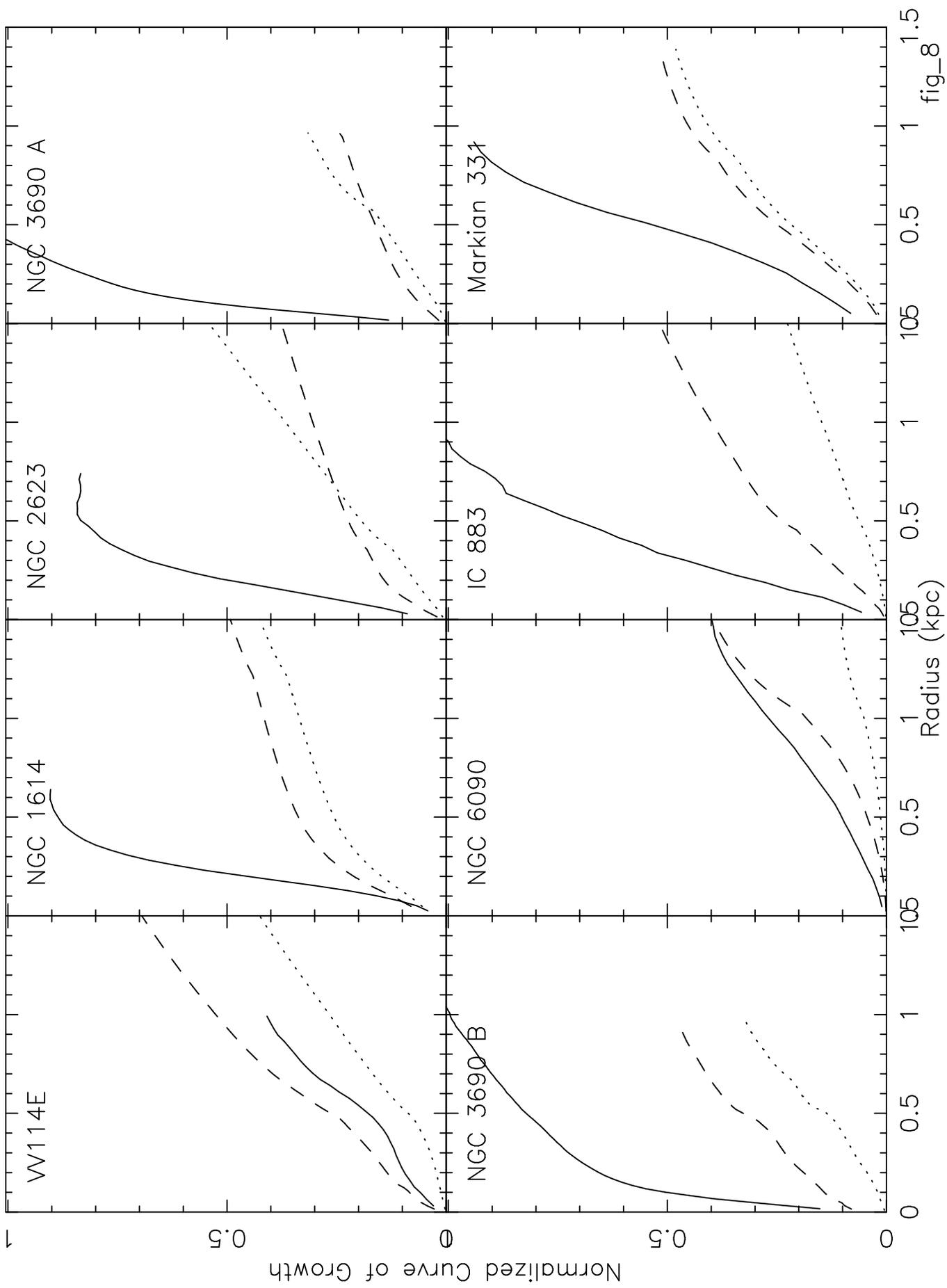
\end{document}